\documentclass[5p,times,twocolumn]{elsarticle}

\usepackage{graphicx}
\usepackage{amssymb}
\usepackage{amsthm}

\usepackage{lineno}

\usepackage{hyperref}
\usepackage{amsmath}
\journal{Nuclear Instruments and Methods in Physics Research Section A}

\begin{document}

\begin{frontmatter}



\title{Radiation Hardness Tests of SiPMs for the JLab Hall~D Barrel Calorimeter\tnoteref{t1}}


\author[jlab]{Yi~Qiang\corref{cor1}}
\ead{yqiang@jlab.org}

\author[jlab]{Carl~Zorn}

\author[jlab]{Fernando~Barbosa}

\author[jlab]{Elton~Smith}

\tnotetext[t1]{Authored by Jefferson Science Associates, LLC under U.S. DOE Contract No. DE-AC05-06OR23177. The U.S. Government retains a non-exclusive, paid-up, irrevocable, world-wide license to publish or reproduce this manuscript for U.S. Government purposes.}
\cortext[cor1]{Corresponding author}

\address[jlab]{Jefferson Lab, 12000 Jefferson Ave, Newport News, VA 23606}

\begin{abstract}
We report on the measurement of the neutron radiation hardness of silicon photomultipliers (SiPMs) manufactured by Hamamatsu Corporation in Japan and SensL in Ireland.
Samples from both companies were irradiated by neutrons created by a 1~GeV electron beam hitting a thin lead target at Jefferson Lab Hall~A.
More tests regarding the temperature dependence of the neutron radiation damage and self-annealing were performed on Hamamatsu SiPMs using a calibrated Am-Be neutron source from the Jefferson Lab Radiation Control group.
As the result of irradiation both dark current and dark rate increase linearly as a function of the 1~MeV equivalent neutron fluence and a temperature dependent self-annealing effect is observed.
\end{abstract}


\end{frontmatter}

\linenumbers

\section{Introduction}

A Silicon photomultiplier (SiPM) is a photon-counting device consisting of multiple avalanche photodiode (APD) pixels operating in Geiger mode.
It is also known as Multi-Pixel Photon Counter (MPPC).
Each APD pixel of the SiPM outputs a pulse signal when it detects photons and the signal output from the SiPM is the total sum of the signals from all APD pixels.
The SiPM offers the high performance needed in photon counting and is used in diverse applications for detecting extremely weak light intensities at the photon-counting level.

For Hall~D At Jefferson Lab~\cite{halld}, we investigated the use of SiPMs to collect light from the Barrel Calorimeter.
One of the requirements for such devices is sufficient radiation hardness to withstand many years of operation.
As the neutron background is expected to be the major source of radiation damage in the Hall~\footnote{An early irradiation test on SiPMs using a series of high activity Cs-137 sources in Jefferson Lab showed that SiPMs are insensitive to electromagnetic radiation and there was no significant change in performance of SiPMs up to 2 krads of gamma irradiation.}, we did a series of tests of the neutron radiation damage to SiPMs at various conditions to evaluate the life time of SiPMs in Hall~D.

\section{Radiation Damage in Silicon Detectors}

The bulk damage in silicon detectors caused by hadrons or higher energy leptons and photons is primarily due to displacement of primary knock-on atoms from the lattice~\cite{Lindstrom2003}.
For neutrons or electrons with kinetic energy above 175~eV and 260~keV, respectively, they will start to generate Frenkel pairs (a pair of a silicon interstitial and a vacancy) along their tracks in silicon material.
With higher energy, more than 35~keV for neutrons and 8 MeV for electrons, a dense cluster of defects will be formed at the end of the primary PKA track.

%Such single displacements resulting in a pair of a silicon interstitial and a vacancy (Frenkel pair) can be generated by e.g. neutrons or electrons with an energy above 175~eV and 260~keV, respectively.
%However, for recoil energies above about 5~keV, a dense agglomeration of defects is formed at the end of the primary PKA track.
%The kinematic lower limits for the production of clusters are 35~keV for neutrons and 8 MeV for electrons.

\begin{figure}[h]
  \begin{center}\includegraphics[width=1.0\linewidth]{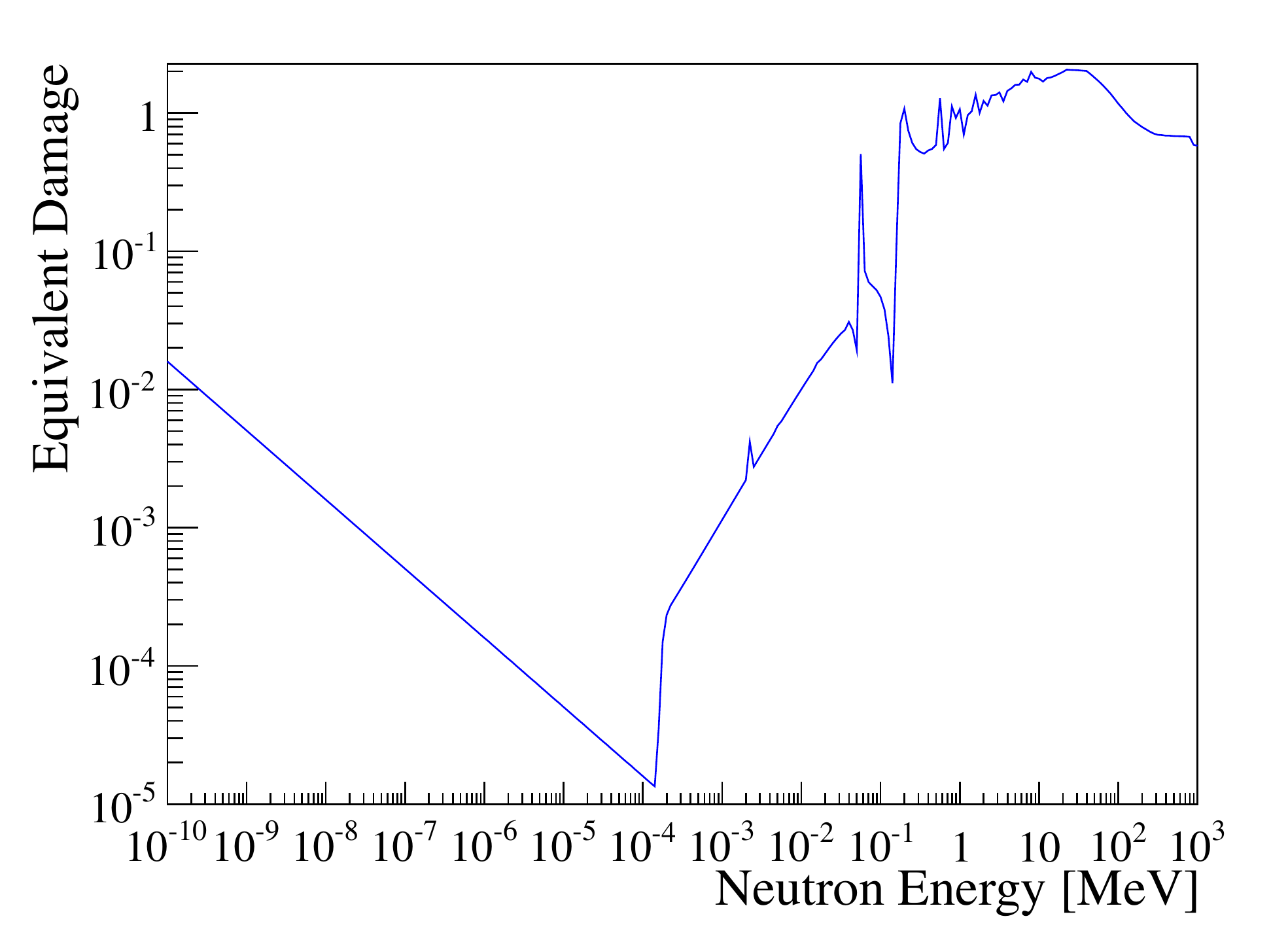}\end{center}
  \caption{Effective damage to Silicon detectors relative to 1~MeV neutron. Data at different energy ranges are taken from References~\cite{Griffin1993,Konobeyev1992,Huhtinen1993}.}\label{fig:neutron_niel}
\end{figure}
The defects will effect silicon detector's performance in various levels depending on their concentration, energy level and the respective electron and hole capture cross-section.
For instance, defects with energy levels in the middle of the forbidden gap acting as recombination/generation centers are responsible for an increase of the reverse current.
Interactions with dopants change the effective doping concentration and therefore change the operating voltage of the detector.
Finally, defects acting as trapping centers reduce the charge collection efficiency.
%Both point defects and clusters can have severe effects on the detector's performance, depending on their concentration, energy level and the respective electron and hole capture cross-section.
%Defects with deep energy levels in the middle of the forbidden gap could act as recombination/generation centers and are hence responsible for an increase of the reverse detector current.
%The removal of dopants by formation of complex defects as well as the generation of charged centers changes the effective doping concentration and the needed operating voltage to fully deplete the detector thickness.
%Finally, such defects could also act as trapping centers affecting the charge collection efficiency.

The radiation damage of neutrons to Silicon detectors has been extensively summarized in the literature.
In this paper, we took data from References~\cite{Griffin1993,Konobeyev1992,Huhtinen1993} to convert the neutron flux to the \textbf{1~MeV neutron fluence} based on the effective damage caused by neutrons with different energies.
The weight of such a conversion is plotted in Fig.~\ref{fig:neutron_niel}.

\section{Test with Electron Beam}

In addition to the goal of testing the neutron radiation damage to SiPMs, we also confirmed our knowledge of the neutron background.

\subsection{Test Setup}

\begin{figure}[h]
  \begin{center}\includegraphics[width=1.0\linewidth]{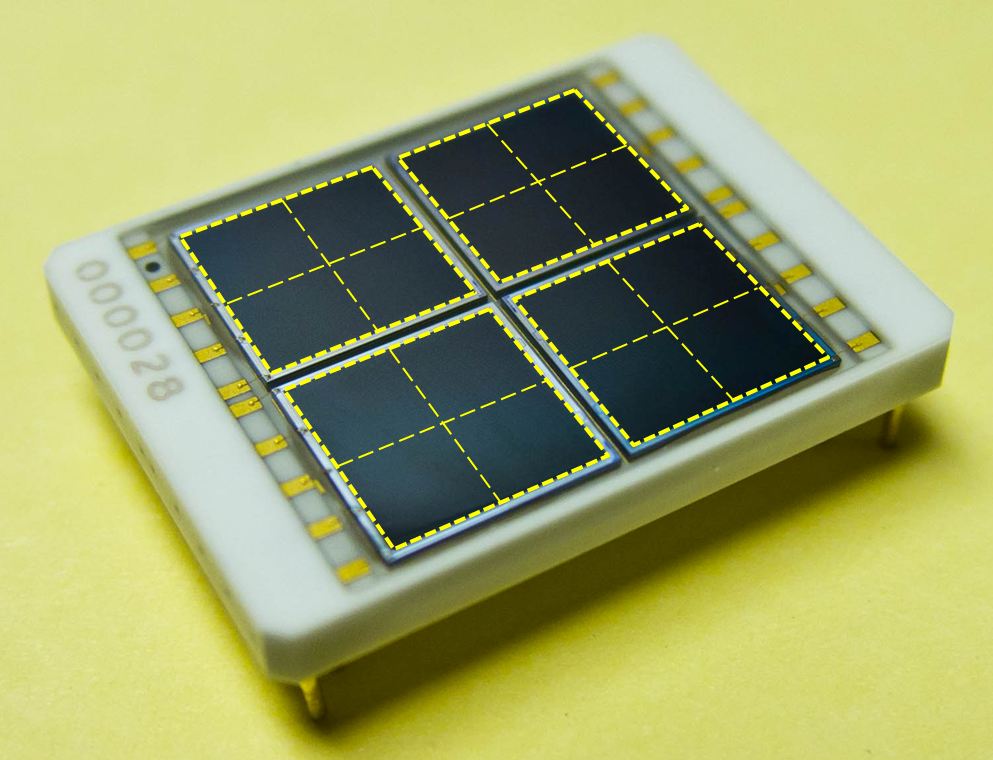}\end{center}
  \caption{A $4\times4$ SiPM array of $3\times3$~mm SiPM cells from Hamamatsu and individual cells are marked by dashed lines.}\label{fig:sipm}
\end{figure}
The irradiation was carried out during the PRex experiment~\cite{prex} at Jefferson Lab Hall~A, with a 1 GeV electron beam incident on a 0.5 mm Pb target.
During the two-day test, one Hamamatsu unit\footnote{A preproduction unit of S10943-0258(X) MPPC with 50~$\mu$m pixels, equivalent to new S12045.} and one SensL unit\footnote{A SPMArray unit based on ceramic design with 35~$\mu$m pixels - Wafer Batch Code X4151-05 using SPM3035 design.} were irradiated.
Both units are $4\times4$ arrays of $3\times3$~mm$^2$ SiPMs and were powered to a gain of 0.75$\times$10$^6$.
A photograph of the Hamamatsu SiPM array is shown in Fig.~\ref{fig:sipm}.

\begin{figure}[h]
  \begin{center}\includegraphics[width=0.9\linewidth]{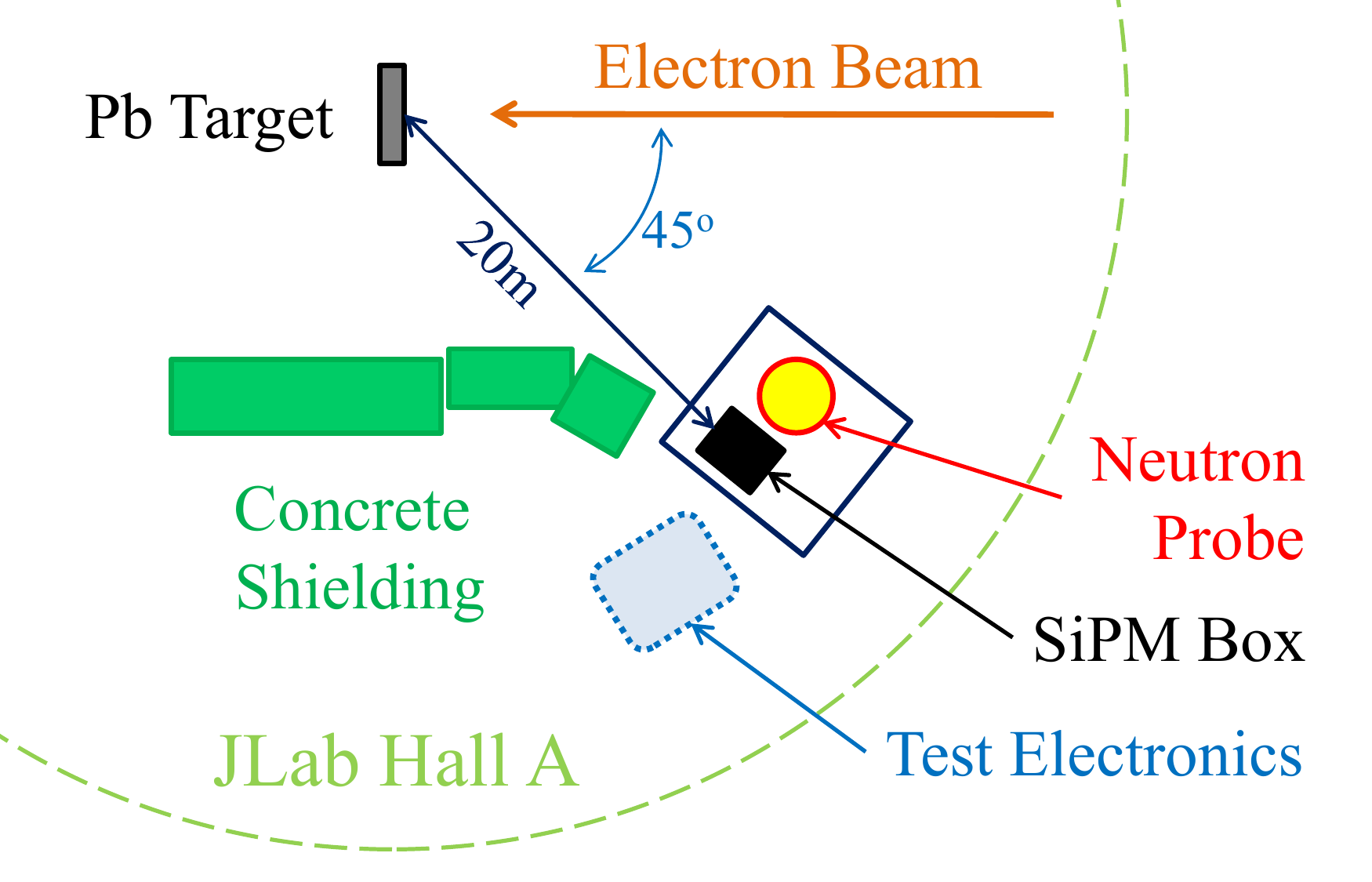}\\
  \includegraphics[width=0.95\linewidth]{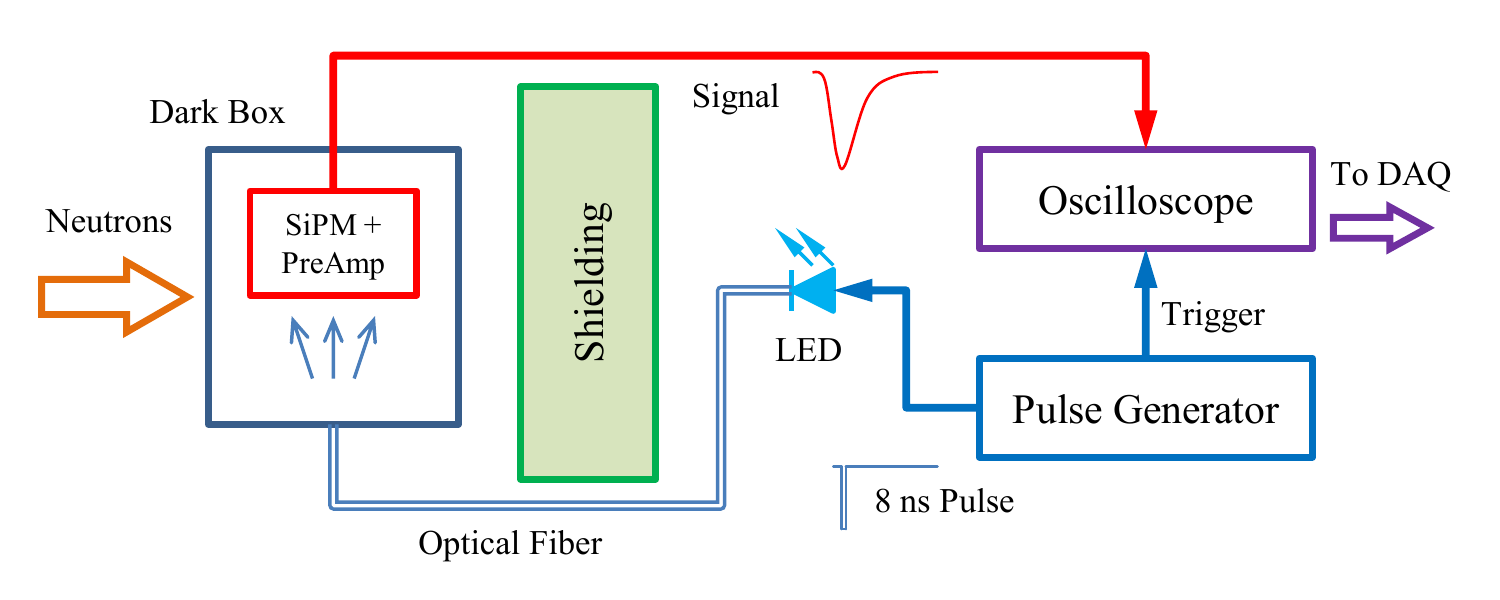}
  \end{center}
  \caption{The floor plan and readout scheme for the SiPM radiation test in Hall~A.}\label{fig:halla_layout}
\end{figure}
Both SiPM units were placed inside a dark box together with their pre-amplifiers.
The light from a pulsed LED was guided into the box through an optical fiber and then diffused by a diffuser to provide uniform light on both units.
The box was positioned 20 meters away from the Hall~A Pb target and 135 degrees backwards to the beam direction as illustrated in Fig.~\ref{fig:halla_layout} to reduce the effect from other sources of radiation such as photons and charged particles.
While the box had direct view of the target, the rest of the equipment was shielded by a concrete wall.
The shape of the output signals, including amplitude and width, was continuously recorded by an oscilloscope.
In the middle of the irradiation period, the power supplies of both SiPMs were turned off intentionally to see whether the powering condition affects the radiation damage.
After the irradiation, both SiPMs were stored at room temperature, 20-25$^\circ$C, for the self-annealing test.

\begin{figure}[h]
  \begin{center}\includegraphics[width=1.0\linewidth]{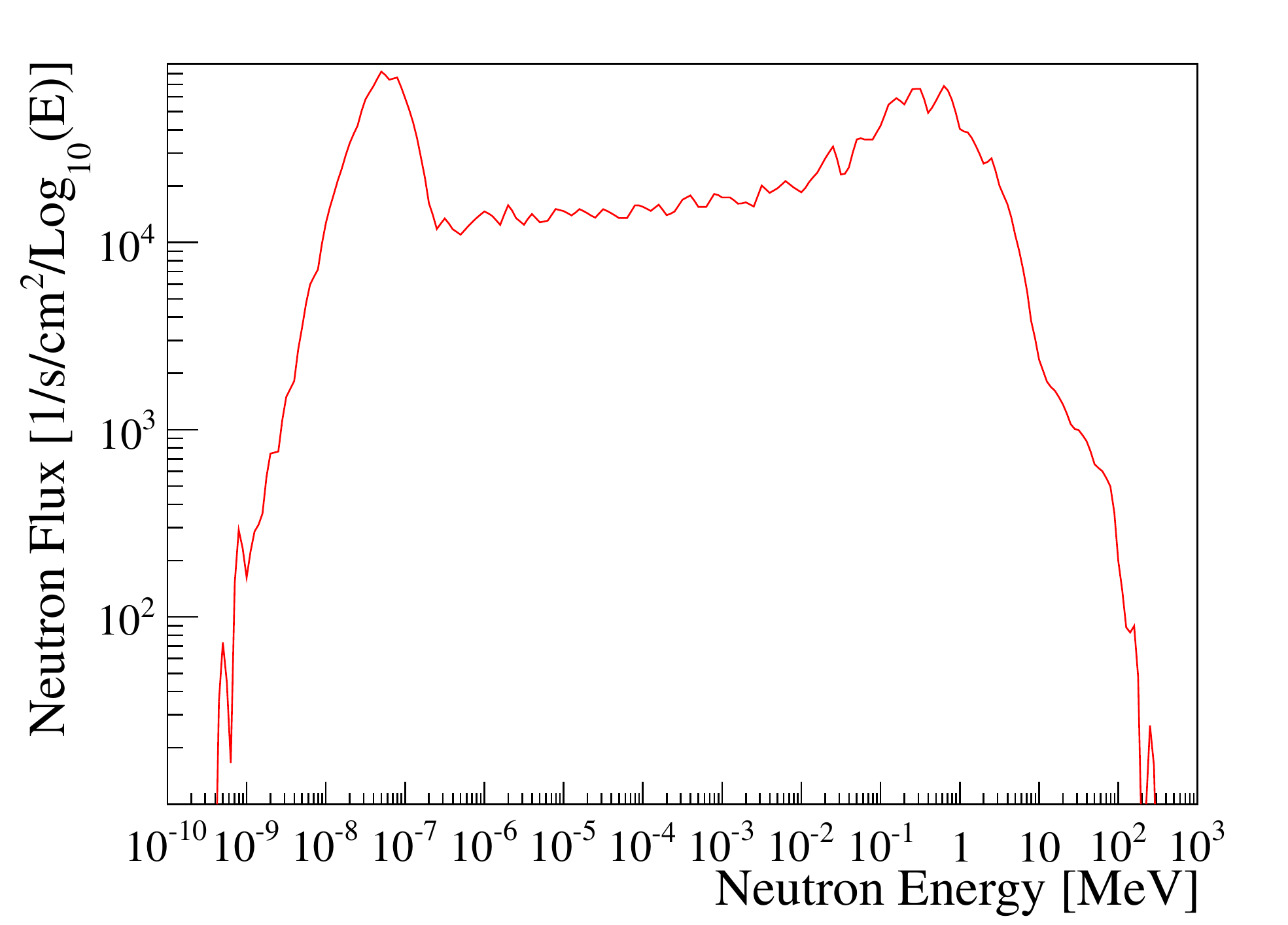}\end{center}
  \caption{The simulated neutron energy spectrum at the SiPM testing area in Hall~A with 50~$\mu$A 1~GeV electron on a 0.5~mm Pb target.}\label{fig:halla_flux}
\end{figure}
A BF$_3$ neutron probe was positioned next to the dark box to monitor the relative change of the neutron flux in real time.
In order to obtain the absolute reading of the effective 1~MeV neutron fluence, a couple of the same type of Hamamatsu SiPMs were later irradiated by a calibrated AmBe neutron source to a similar damage level.
As the AmBe source has a well known narrow energy spectrum peaking at about 4~MeV~\cite{VegaCarrillo2002}, its fluence was calculated by convoluting the neutron flux spectrum with the effective damage curve shown in Fig.~\ref{fig:neutron_niel}.
Then the fluence in Hall~A was calculated by comparing the damage the SiPMs received in both cases.

As a result, we determined that the two SiPMs in Hall~A received a fluence of about $3.7\times10^9~\mathrm{n_{eq}/cm^2}$.
The fluence measurement also provides a good bench mark of our knowledge of the radiation level in the experimental halls.
The neutron flux in Hall~A was simulated in a GEANT3 framework customized to the electron-beam environment~\cite{Degtyarenko2000,Degtyarenko2000a,Degtyarenko2000b}.
The resulting energy spectrum is shown in Fig.~\ref{fig:halla_flux} and the fluence obtained from such a simulation is consistent with the measured value within 50\%.
The same code predicts that the $3.7\times10^9~\mathrm{n_{eq}/cm^2}$ fluence will be reached in about 13 years in Hall~D with its high intensity GlueX running~\footnote{Such a high intensity refers to a Bremsstrahlung photon flux of about 100~MHz/GeV close to the 12~GeV endpoint.} on a 30~cm liquid Hydrogen target.

\subsection{Results}

\subsubsection{Dark Current}

\begin{figure}[h]
  \begin{center}\includegraphics[width=1.0\linewidth]{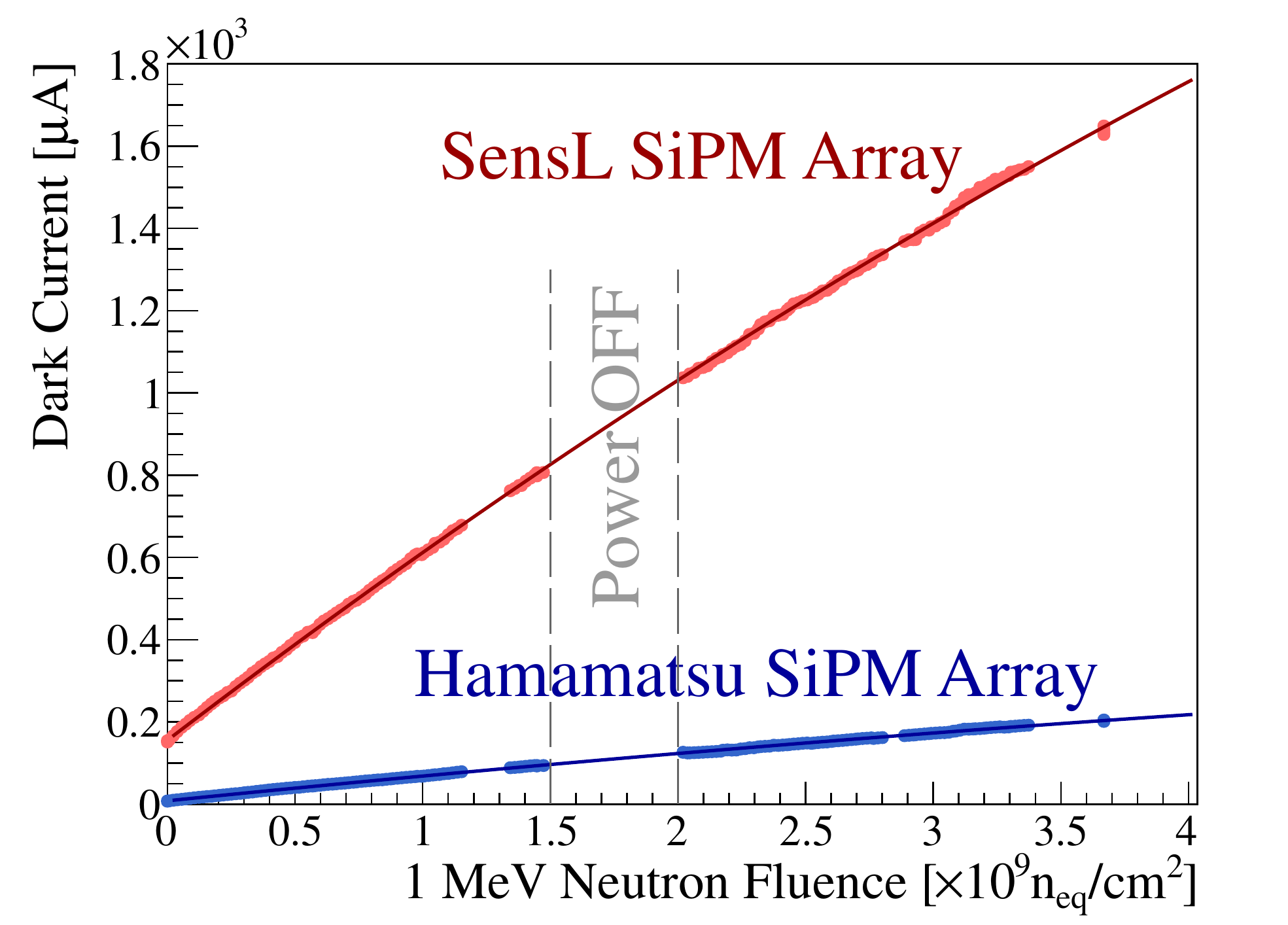}\end{center}
  \caption{(Color online) The increase of the dark current of SiPMs as a function of the neutron fluence during the two-day irradiation in Hall~A.
  The curves are fits using second order polynomials.
  The period with no data marked by grey dash lines corresponds to the time when the sensors were not powered, but continued to be irradiated.}\label{fig:dark_current}
\end{figure}
The change of the SiPM dark current as a function of the accumulated neutron fluence is plotted in Fig.~\ref{fig:dark_current}.
As the beam was turned on, the dark current of both SiPMs started to increase immediately.
By comparing the trends of the damage before and after the period when the power was turned off, one can see that the neutron damage remains the same no matter whether the unit is powered or not.
Over the course of the test, the dark current increased by a factor of about 10 for the SensL SiPM Array, $160~\mathrm{\mu A}\to1.6~\mathrm{mA}$, and 25 for the Hamamatsu SiPM Array, $8~\mathrm{\mu A}\to200~\mathrm{\mu A}$.

\subsubsection{Signal}

\begin{figure}
  \begin{center}\includegraphics[width=1.0\linewidth]{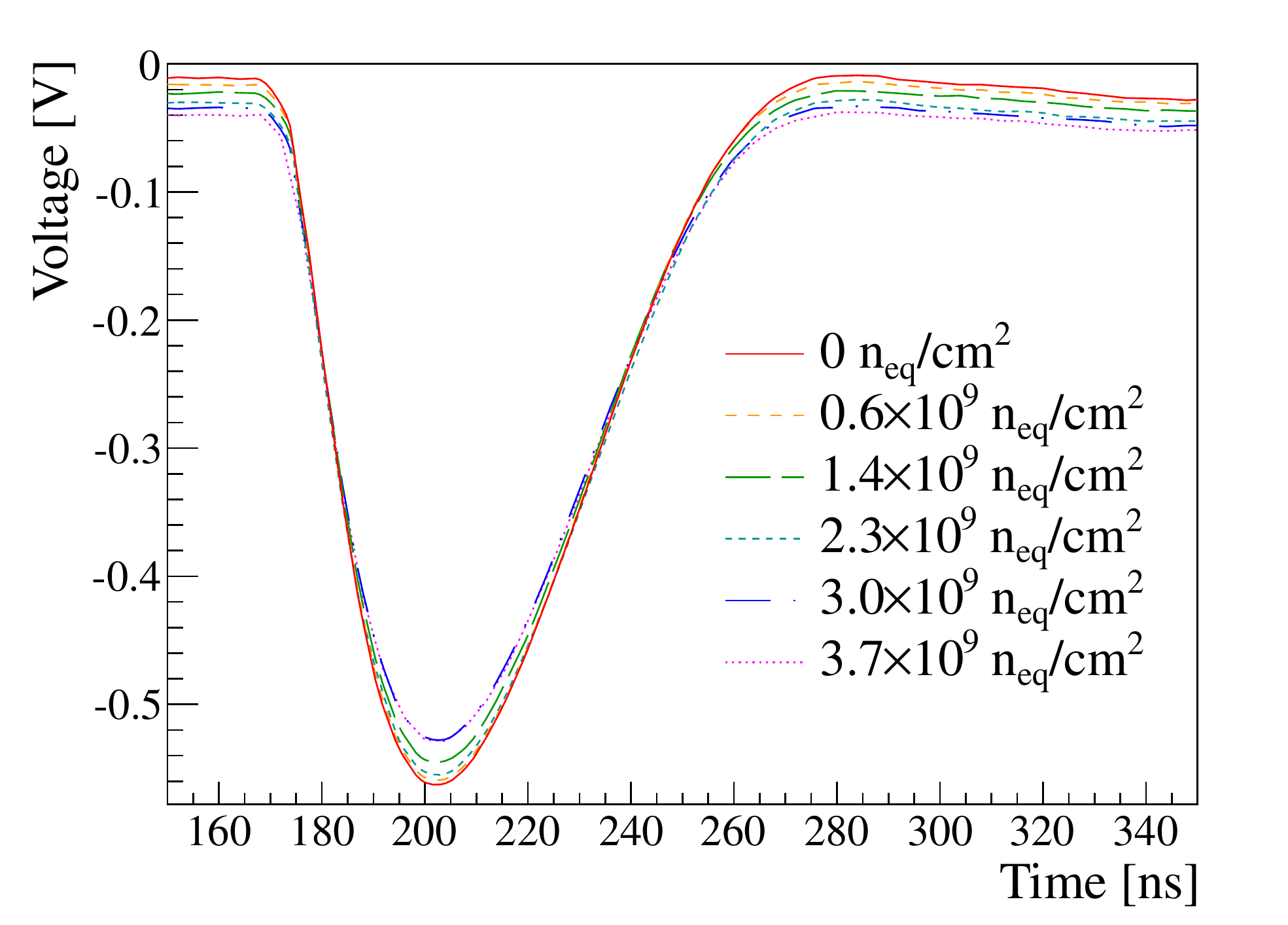}\end{center}
  \caption{(Color online) Signals from the Hamamatsu SiPM array recorded by the oscilloscope during the irradiation.}\label{fig:signal}
\end{figure}
\begin{figure}
  \begin{center}\includegraphics[width=1.0\linewidth]{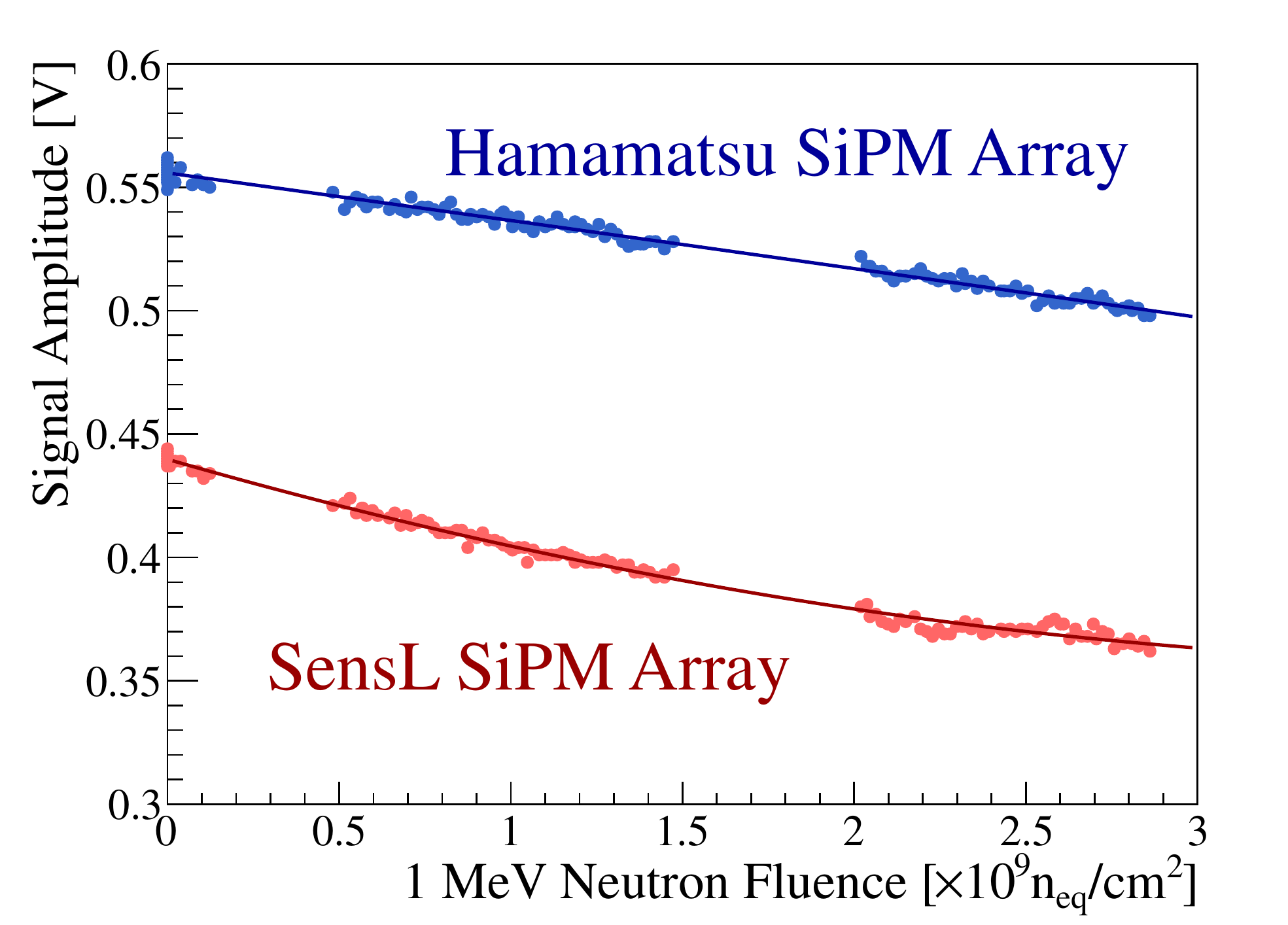}\\
  \includegraphics[width=1.0\linewidth]{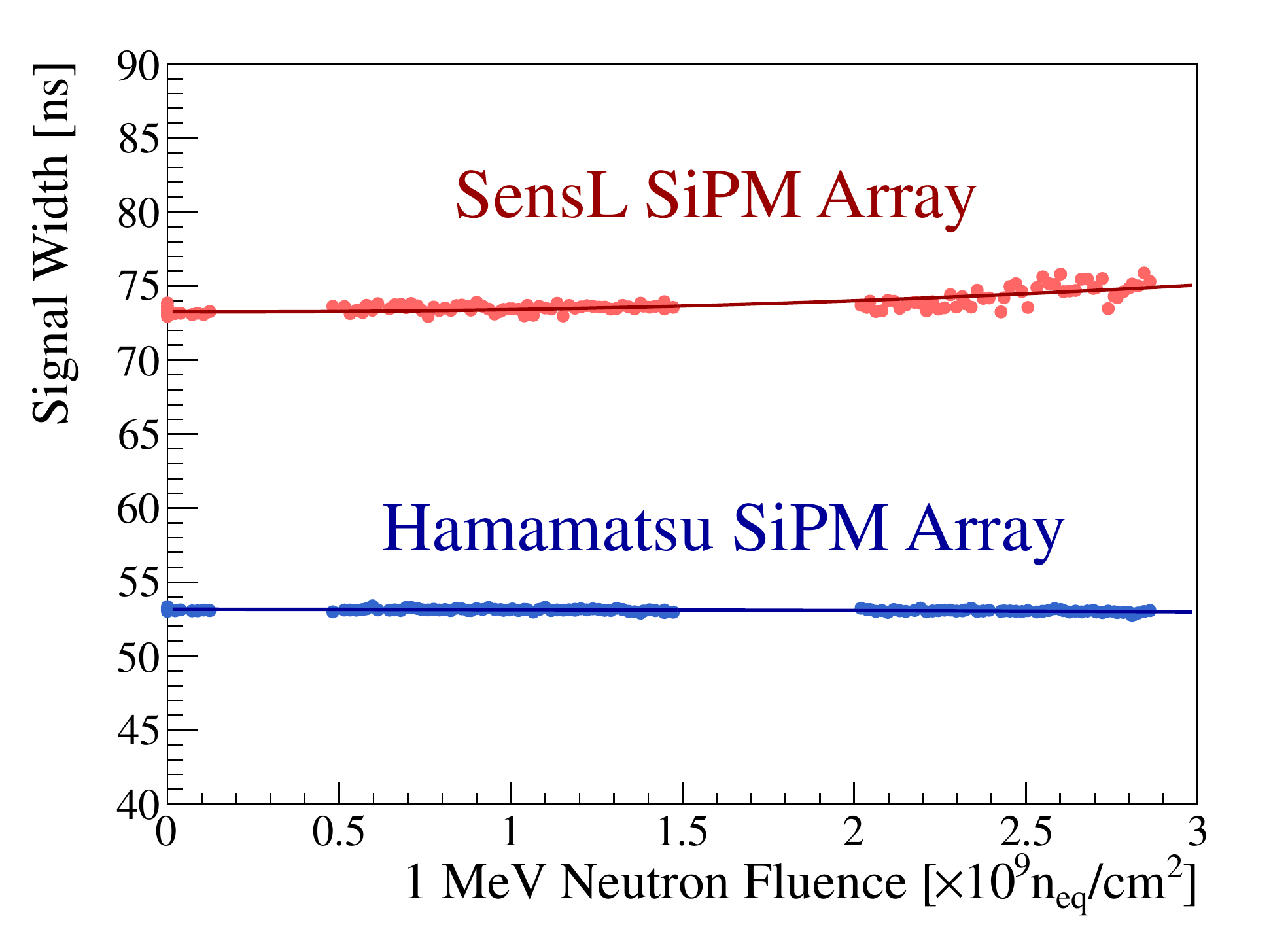}\end{center}
  \caption{(Color online) The impact of neutron radiation damage on the signal shape, amplitude and width, of the SiPM output.
  The amplitude dropped slightly while no change was observed in the width.
  The curves are fits using second order polynomials.}\label{fig:signal_detail}
\end{figure}
The output signals from both SiPMs stayed relatively stable in contrast to the dramatic change of the dark current.
The amplitude and width (50\% to 50\%) are plotted in Fig.~\ref{fig:signal} and \ref{fig:signal_detail}.
While the width shows no noticeable change, the amplitude dropped by about 10\%.

\subsubsection{I-V Curve}

\begin{figure}[h]
  \begin{center}\includegraphics[width=1.0\linewidth]{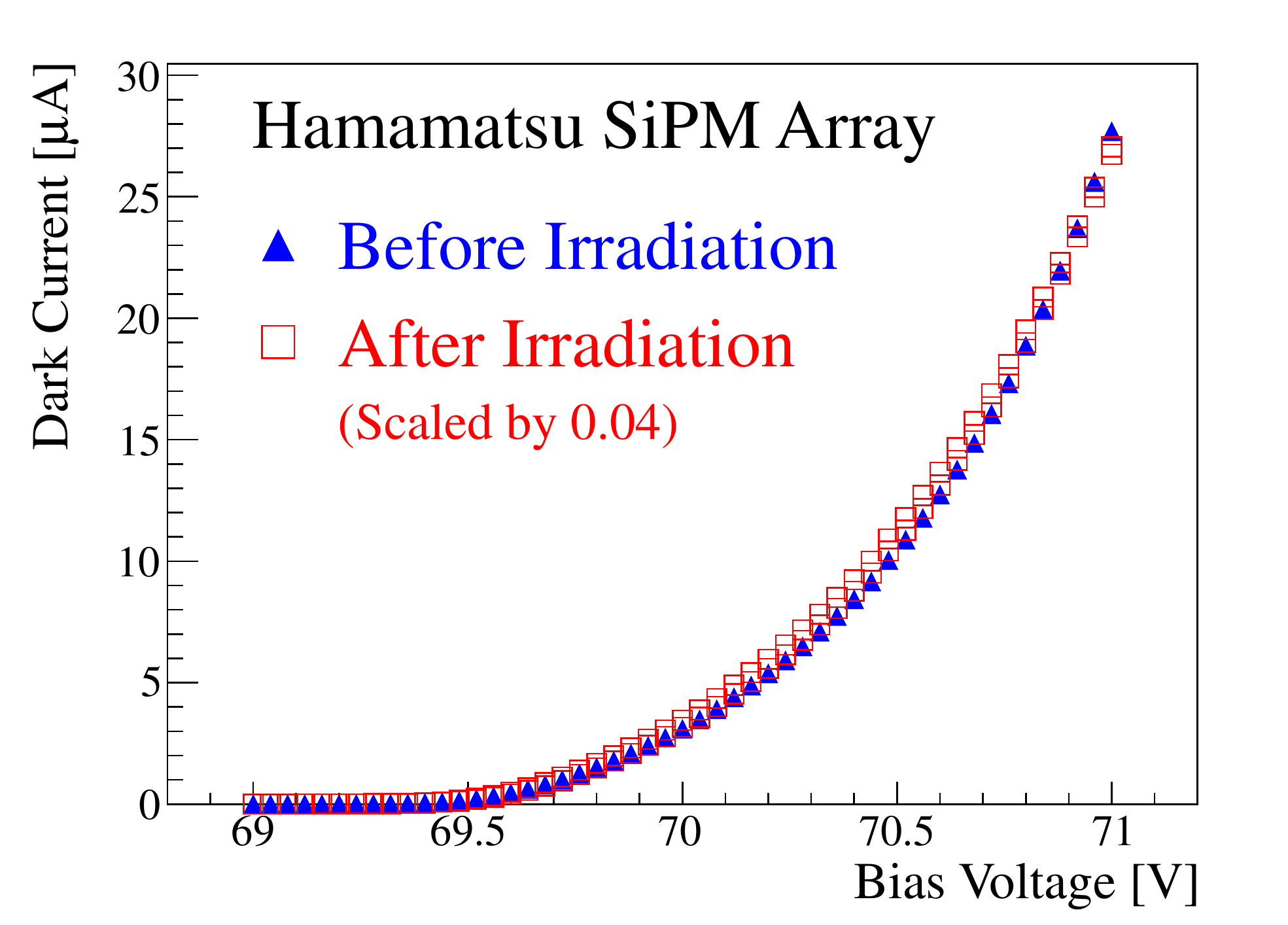}\end{center}
  \caption{(Color online) The I-V curve of Hamamatsu SiPM array before and after the irradiation.
  The dark current after irradiation is scaled by a 0.04 for a better visual comparison and clearly the break down voltage was not effected by the neutron radiation.}\label{fig:iv_curve}
\end{figure}
The current vs. voltage (I-V) curves before and after irradiation are also compared for both units, and the comparison of the Hamamatsu SiPM array is shown in Fig.~\ref{fig:iv_curve}.
Other than an overall change in scale, the I-V curves stay the same for both units and indicate that the break down voltage of the SiPM is not impacted by the neutron radiation.

\subsubsection{Self-Annealing}

\begin{figure}[h]
  \begin{center}\includegraphics[width=1.0\linewidth]{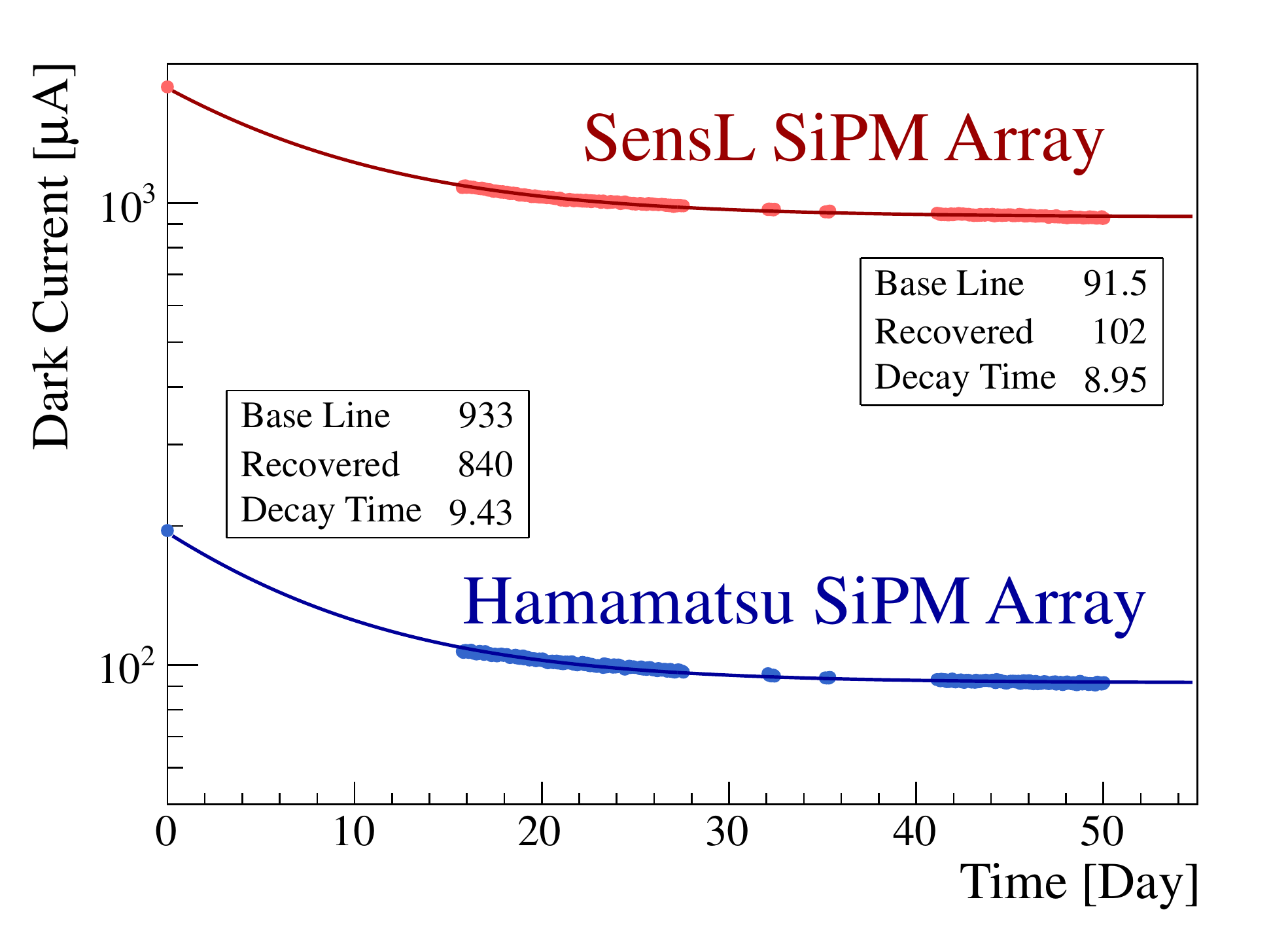}\end{center}
  \caption{(Color online) The decline of dark current during SiPM's self-annealing at room temperature after the irradiation.
  The time constant for both samples is about 10 days and approximately half of the damage recovered.
  The data are fitted by an exponential function described in Eq.~(\ref{eqn:decay}).}\label{fig:annealing_current}
\end{figure}
\begin{figure}[h]
  \begin{center}\includegraphics[width=1.0\linewidth]{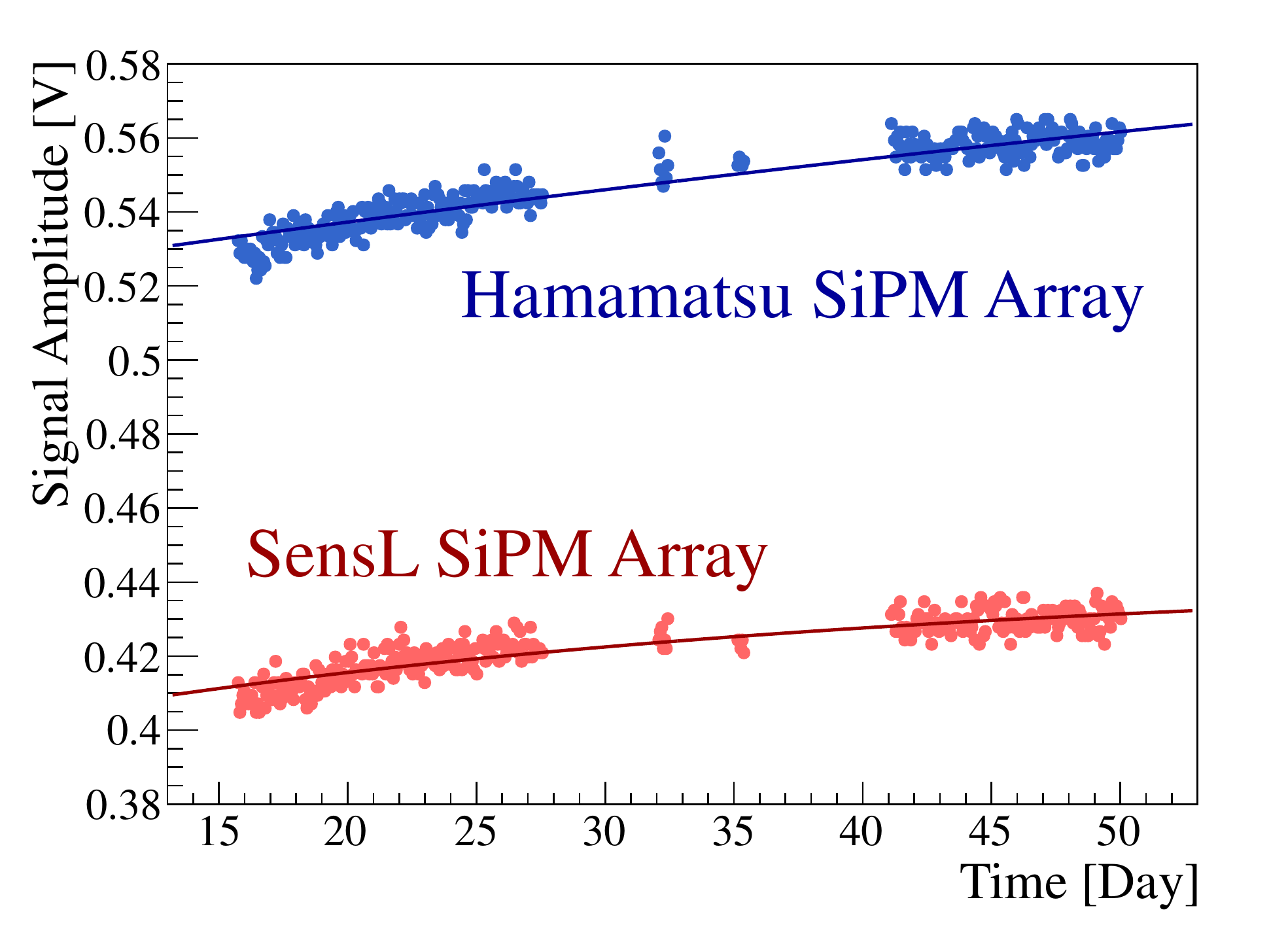}\end{center}
  \caption{(Color online) The recovery of the signal amplitude.
  Signals from both samples returned to their levels before neutron irradiation.}\label{fig:annealing_signal}
\end{figure}
Following the delivery of an intense prompt dose of radiation, one finds that not all damage to the lattice is permanent.
So we studied the response of the sensors as they rested without further irradiation (self-annealing).
The results of the self-annealing at room temperature after the prompt irradiation are plotted in Fig.~\ref{fig:annealing_current} and \ref{fig:annealing_signal}.
Both the dark current and signal amplitude recovered over time with a time constant close to 10 days.
About half of the damage to the dark current recovered and the signal amplitude completely returned to the level before the irradiation.

\section{Temperature Tests with a AmBe Neutron Source}

The Hall~A irradiation test revealed that the lifetime of the SiPM will be marginal in Hall~D given the experimental requirement on the dark rate.
Fortunately, a lower dark rate at a fixed gain can be achieved if the SiPM is cooled.
%By doing so, it will certainly allow more room for the increase of the dark noise caused by the radiation damage.
Cooling the SiPMs to 5$^\circ$C will reduce the dark rate to about 1/3 compared with 20$^{\circ}$C and will certainly allow more room for the increase of the dark rate caused by the neutron radiation damage.
However, whether such a dependence will be effected by radiation damage was unknown, therefore a systematic study of the temperature dependence of the neutron radiation damage and annealing was performed using a calibrated AmBe neutron source.

\subsection{Test Procedure}

Twelve $1\times1$~mm$^2$ SiPM units from Hamamatsu with 50~$\mu$m pixels (PN\# S10362-11-050C) were irradiated during this test.

Since the previous Hall~A irradiation test shows no correlation between the damage and the powering condition, all the SiPMs were not powered during the irradiation or annealing except when their dark currents were measured.
If not specifically mentioned, the dark current was always measured at room temperature regardless of the SiPM's irradiation or annealing temperatures.
This allows direct comparison of results between samples regardless of the temperature at which they were irradiated or annealed.
It is assumed that taking the test samples to room temperature during the short time of the measurement does not significantly influence the results.
The unit was then powered off and put back to its previous temperature after the dark current was measured.
All the SiPMs were powered to a gain of $0.84\times10^6$ during the dark current measurements.
The gain was determined using the ADC spectra and more details can be found in Sec.~\ref{sec:rate}.

\begin{figure}[h]
  \begin{center}\includegraphics[width=0.9\linewidth]{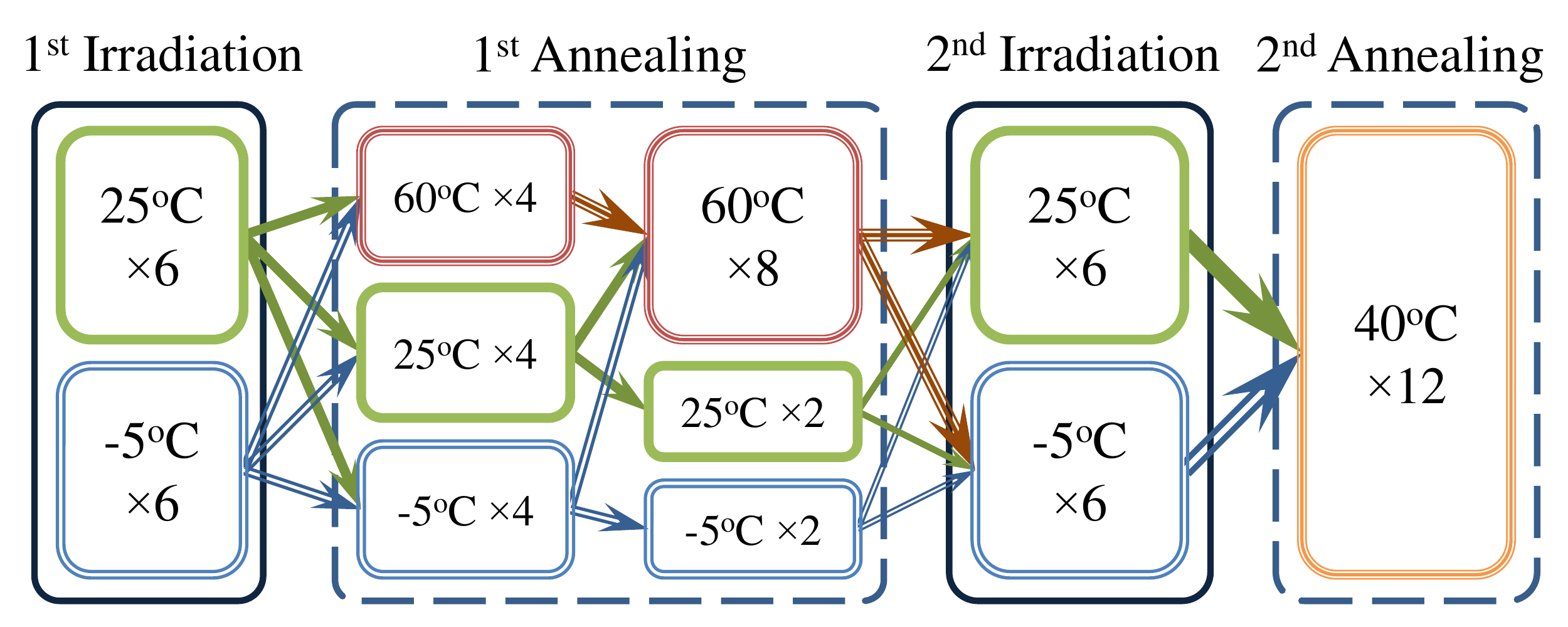}\end{center}
  \caption{(Color online) Steps of the irradiation/annealing temperature dependence test.
  Numbers in boxes indicate number of samples tested in each group.}\label{fig:test_steps}
\end{figure}
The test consists of the following steps as shown in Fig.~\ref{fig:test_steps}:
\begin{enumerate}[1.]
  \item In the first stage, six units were irradiated at $-5^\circ$C while the remaining six units were irradiated at room temperature, $\sim25^\circ$C.
      The irradiation lasted four days and the total fluence each unit received was about $1.4\times10^{9}~\mathrm{n_{eq}/cm^2}$.
  \item Right after the first irradiation, all the units were immediately stored at three different temperatures, $-5^\circ$C, 25$^\circ$C and 60$^\circ$C, for their first annealing.
      Every group had two units from each temperature group of the first irradiation.
      With all the units annealed, half of the units from the $-5^\circ$C and 25$^\circ$C annealing groups were further heated to 60$^\circ$C to investigate any additional recovery while the rest of the units were still kept at their original temperatures.
  \item After the first annealing was completed, all the units were irradiated again at $-5^\circ$C or 25$^\circ$C for four more days with an additional fluence of $1.7\times10^{9}~\mathrm{n_{eq}/cm^2}$, to see whether the radiation damage due to the first irradiation would effect the subsequent damage rate.
  \item At the end, all the units were heated to 40$^\circ$C for a final accelerated annealing.
\end{enumerate}

\subsection{Results}

\subsubsection{Temperature Dependence of Radiation Damage and Recovery}

The average current of all units before the irradiation test is $86\pm3$~nA and the uncertainty is the standard deviation of the measurements of individual units.
After the first irradiation, the average current of the group at $-5^\circ$C increased to $771\pm66$~nA, and for the 25$^\circ$C group, the current went up to $660\pm38$~nA.
Such a 110~nA difference suggests two possibilities, one is a temperature dependence of the radiation damage and the other one is a temperature dependence of the damage recovery.

For the units annealed at 60$^\circ$C during the first annealing, the average dark currents of the units from the $-5^\circ$C and 25$^\circ$C irradiation groups dropped to $365\pm61$~nA and $341\pm23$~nA, respectively.
The consistency of these two values excludes the temperature dependence of the radiation damage to the limit of the variation among samples.

On the other hand, if the recovery at $-5^\circ$C is much weaker or slower than 25$^\circ$C, the recovery during the four days of irradiation will reduce the damage to the 25$^\circ$C group more.
The results of the annealing at different temperatures indeed confirm this hypothesis and the recovery at higher temperature is faster and stronger.

\begin{figure}[h]
  \begin{center}\includegraphics[width=1.0\linewidth]{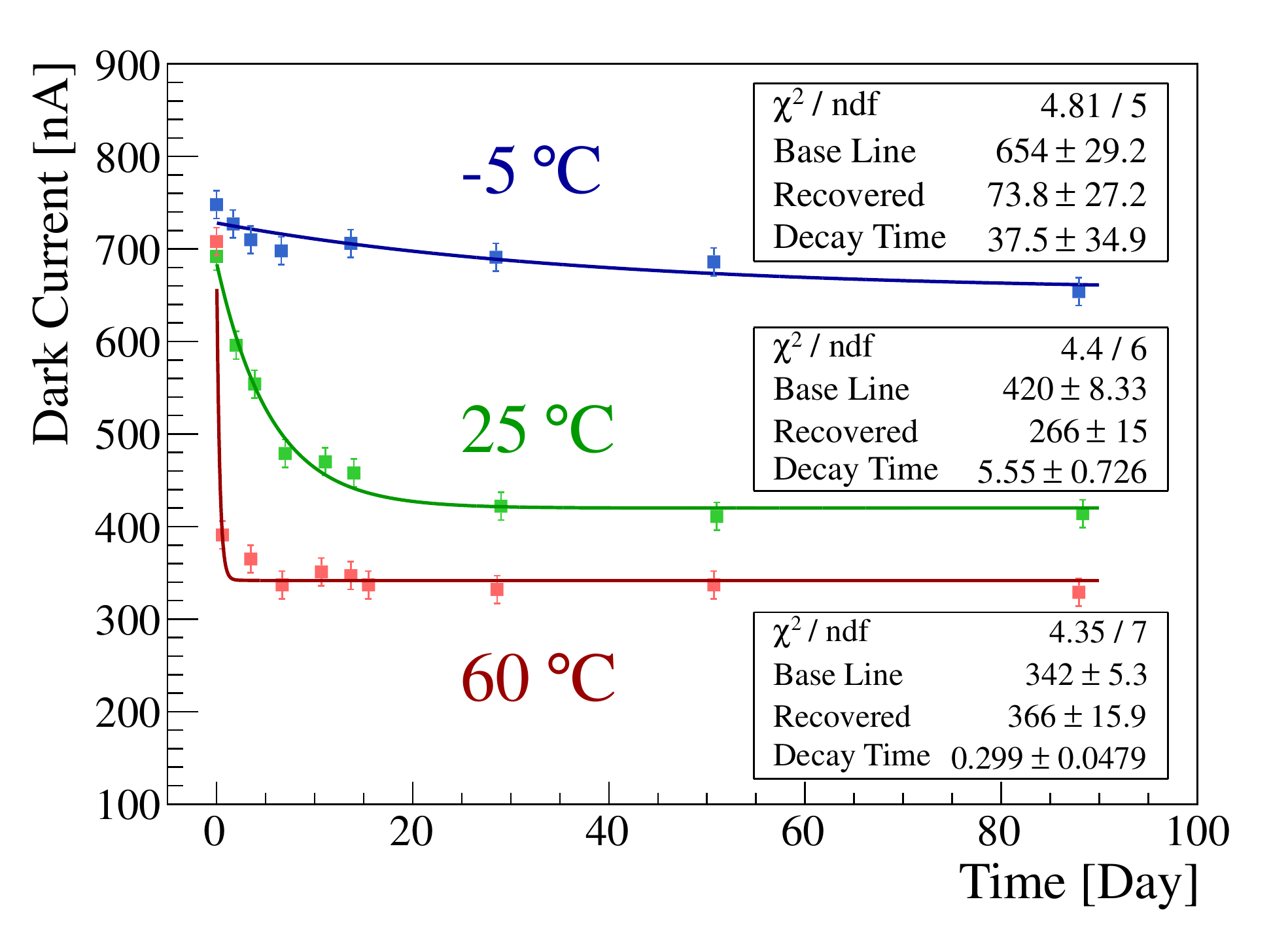}\end{center}
  \caption{(Color online) The recovery of the dark current at different temperatures after the first irradiation.
  All the values were measured at 25$^\circ$C and the uncertainties only include the accuracy of the measurements.}\label{fig:annealing_all}
\end{figure}
\begin{figure}[h]
  \begin{center}\includegraphics[width=1.0\linewidth]{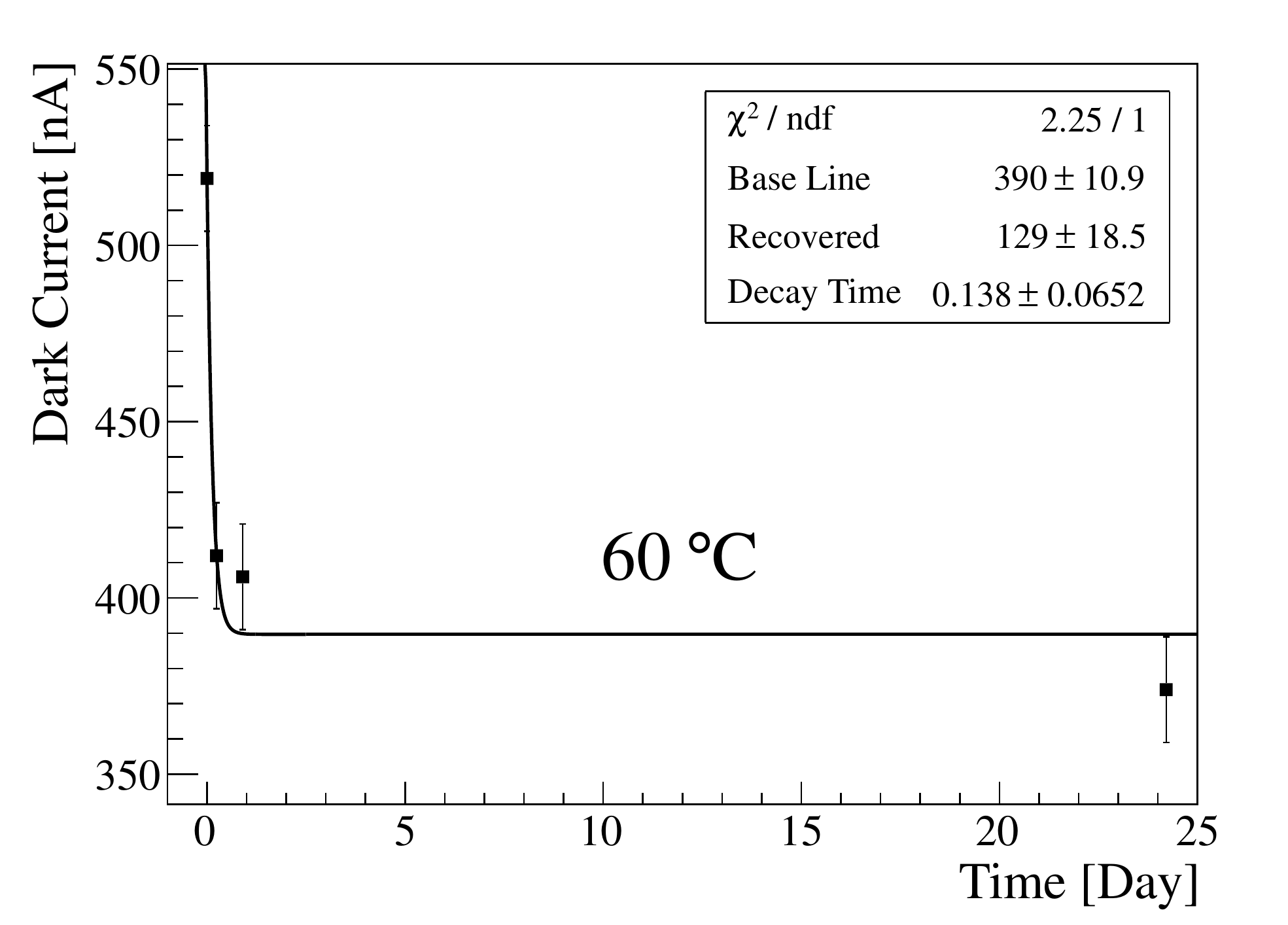}\end{center}
  \caption{The secondary recovery at 60$^\circ$C at the end of the first annealing of selected units previously annealed at $-5^\circ$C or 25$^\circ$C.}\label{fig:annealing_60}
\end{figure}
\begin{figure}[h]
  \begin{center}\includegraphics[width=1.0\linewidth]{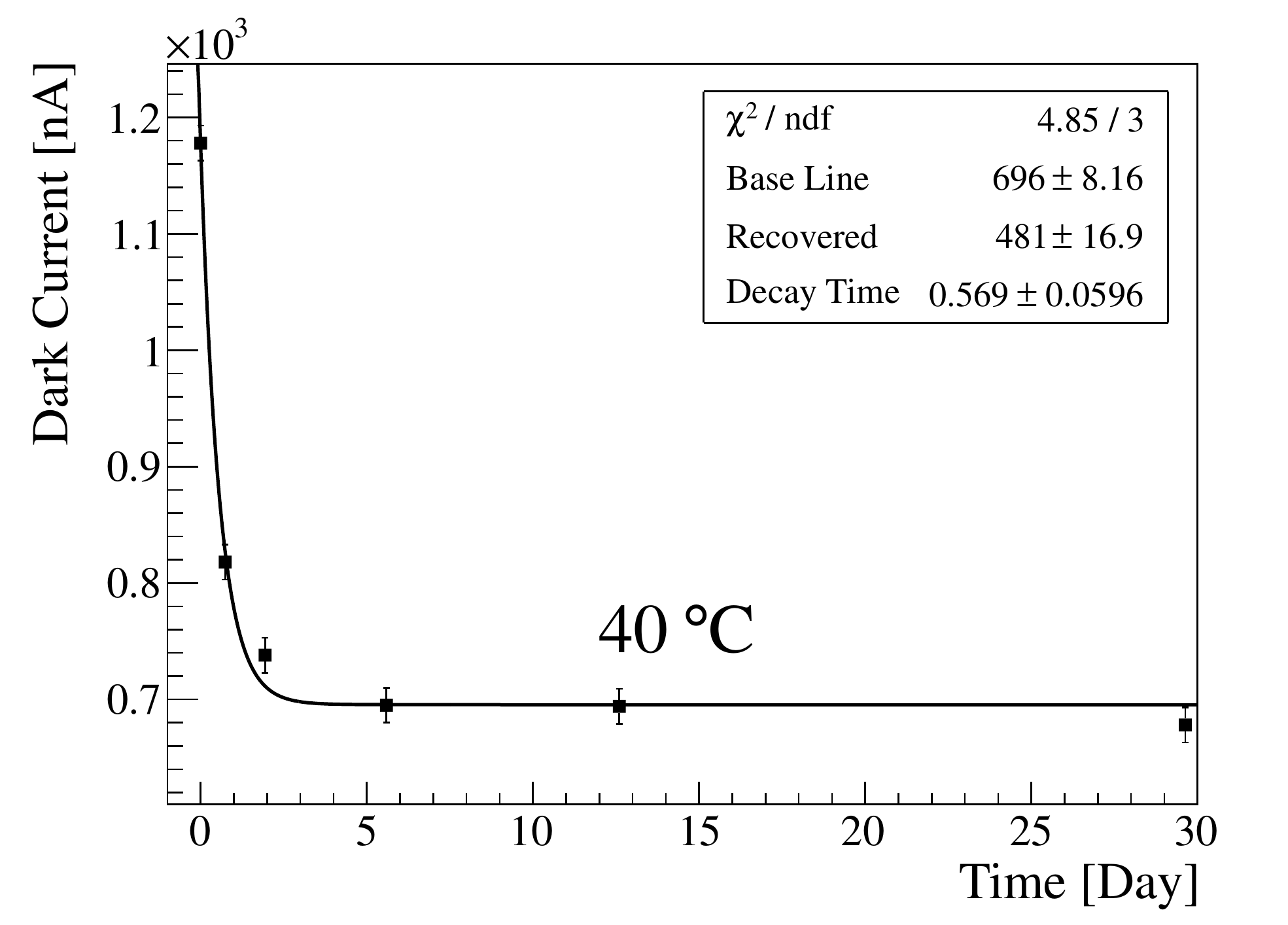}\end{center}
  \caption{The final recovery of SiPMs at 40$^\circ$C after the second irradiation.}\label{fig:annealing_40}
\end{figure}
The recovery curves at various temperatures are shown in Fig.~\ref{fig:annealing_all}, \ref{fig:annealing_60} and \ref{fig:annealing_40}.
The uncertainties in these plots only include the estimated uncertainty of the current measurement, 15 nA, while the variation among individual units is not included since it has no impact on the fit of the recovery time constant.
The data are fitted with a simple exponential decay function with a constant baseline offset:
\begin{equation}\label{eqn:decay}
    I = b + a\cdot e^{-t/\tau}
\end{equation}
where $b$ is the baseline dark current after annealing, $a$ is the recoverable damage and $\tau$ is the time constant.

Fig.~\ref{fig:annealing_all} shows the recovery of the SiPM dark current at $-5^\circ$C, 25$^\circ$C and 60$^\circ$C right after the first irradiation.
It is very clear that the units annealed at higher temperature recover faster and reach a lower asymptotic dark current.
In order to see whether the baseline will be fixed after first annealing at a lower temperature, half of the units from the groups of $-5^\circ$C and 25$^\circ$C were heated to 60$^\circ$C for a secondary annealing.
As shown in Fig.~\ref{fig:annealing_60}, additional recovery was observed and the time constant is consistent with the one obtained from the original 60$^\circ$C group.
The time constant of 40$^\circ$C annealing were measured later after the second irradiation, as shown in Fig.~\ref{fig:annealing_40}.

\begin{figure}[h]
  \begin{center}\includegraphics[width=1.0\linewidth]{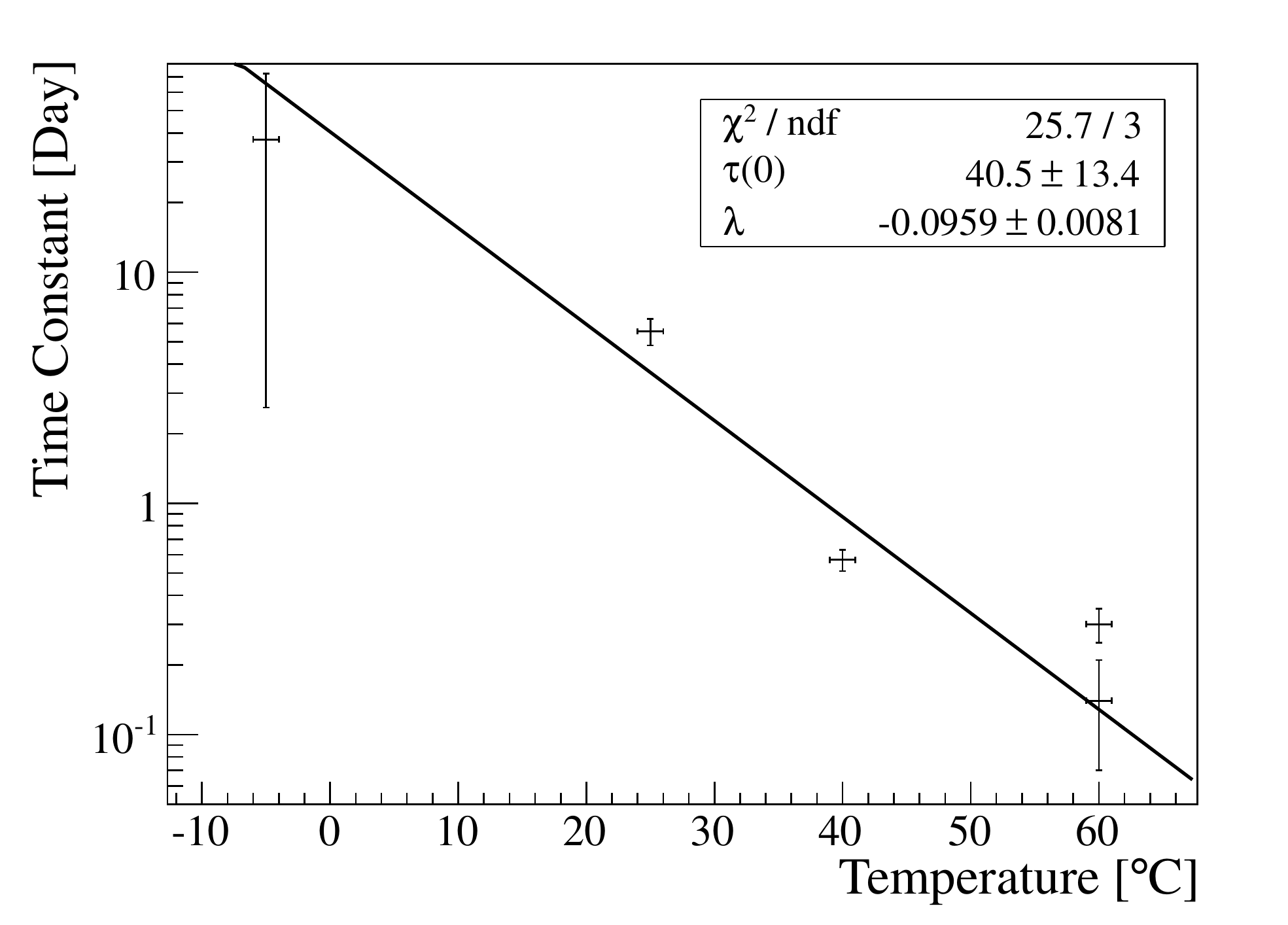}\end{center}
  \caption{The dependence of the time constant $\tau$ in Eq.~(\ref{eqn:decay}) of the annealing on the temperature.
  Such a dependence is fitted by an exponential curve.}\label{fig:time_temp}
\end{figure}
The temperature dependence of the recovery time constants can be described by an exponential curve
\begin{equation}
    \tau(T) = 41\cdot e^{-0.10\cdot T}\mathrm{day}
\end{equation}
as plotted in Fig.~\ref{fig:time_temp}.

\begin{figure}[h]
  \begin{center}\includegraphics[width=1.0\linewidth]{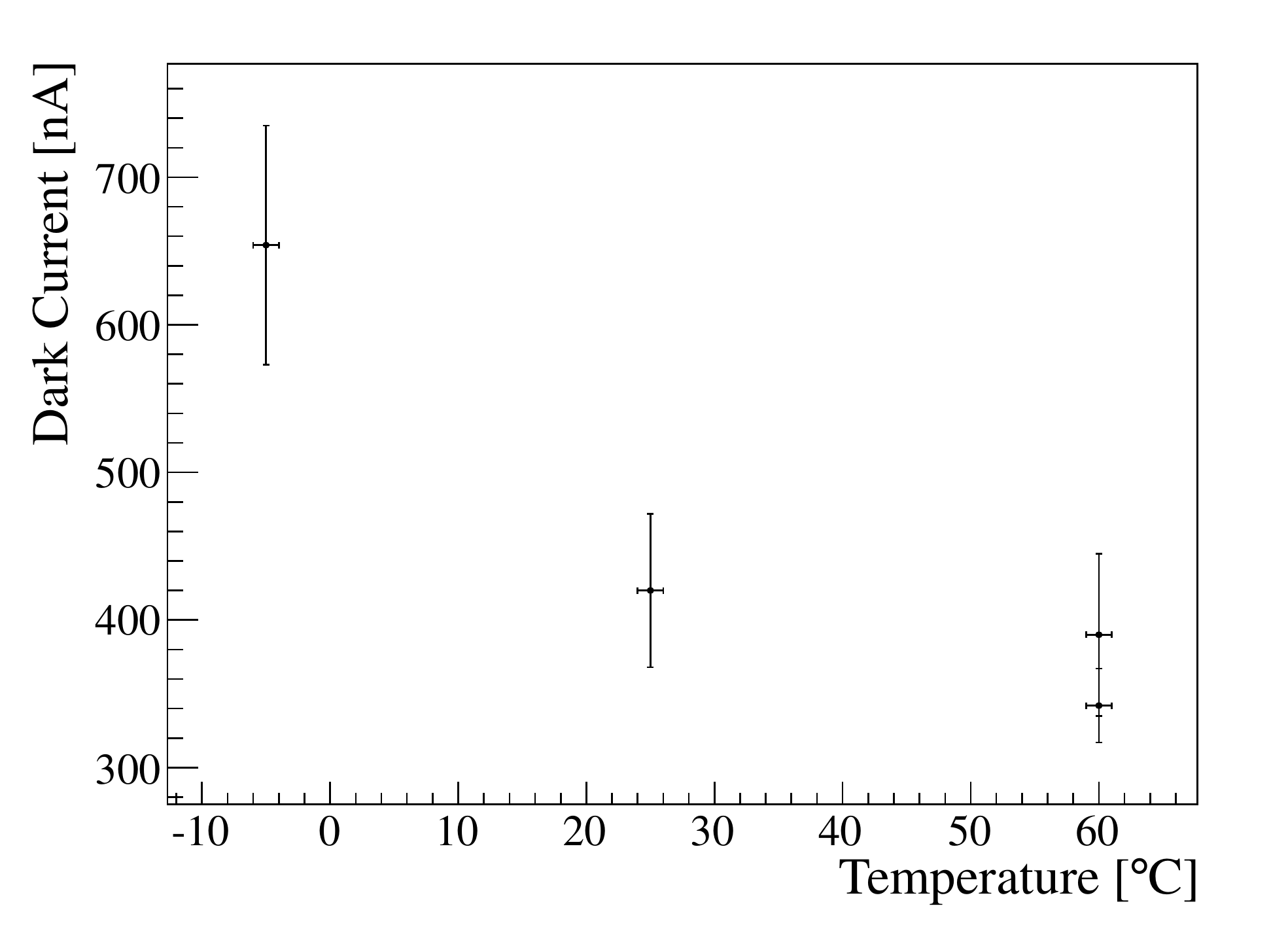}\end{center}
  \caption{The dependence of the baseline $b$ in Eq.~(\ref{eqn:decay}) of the annealing on the temperature.}\label{fig:baseline_temp}
\end{figure}
Fig.~\ref{fig:baseline_temp} shows the temperature dependence of the baseline.
The uncertainties in the plot now include the variations from individual units.
Given the limited accuracy and the number of data points, the function of the temperature dependence can not be well determined.
%Given the accuracy of the data, it is hard to pin down the dependence function, and the second order polynomial gives a slightly better fit than the others.
It is clear nevertheless that when the annealing temperature is above 40$^\circ$C, the change of the baseline can not be clearly identified given the variation of individual units.
%The baseline of the annealing at 40$^\circ$C after the first irradiation is estimated to be $363\pm47$~nA.

As already discussed, the damage will further recover if higher temperature is applied later.
In order to see whether the recovery would reverse when the annealed units are stored at a lower temperature, half of the units annealed at 60$^\circ$C were put into a freezer for several weeks, and no indication of any increase in the dark current was found.

%Another interesting fact is found as following.
The two units which were always kept at $-5^\circ$C during the annealing after the first irradiation were annealed at 40$^\circ$C with the rest of the units after the second irradiation.
At the end, both units recovered to a level consistent with all the other units.
This fact suggests that the temporary damage resulting from previous irradiations can always be recovered with sufficiently high annealing temperature.

\begin{figure}[h]
  \begin{center}\includegraphics[width=1.0\linewidth]{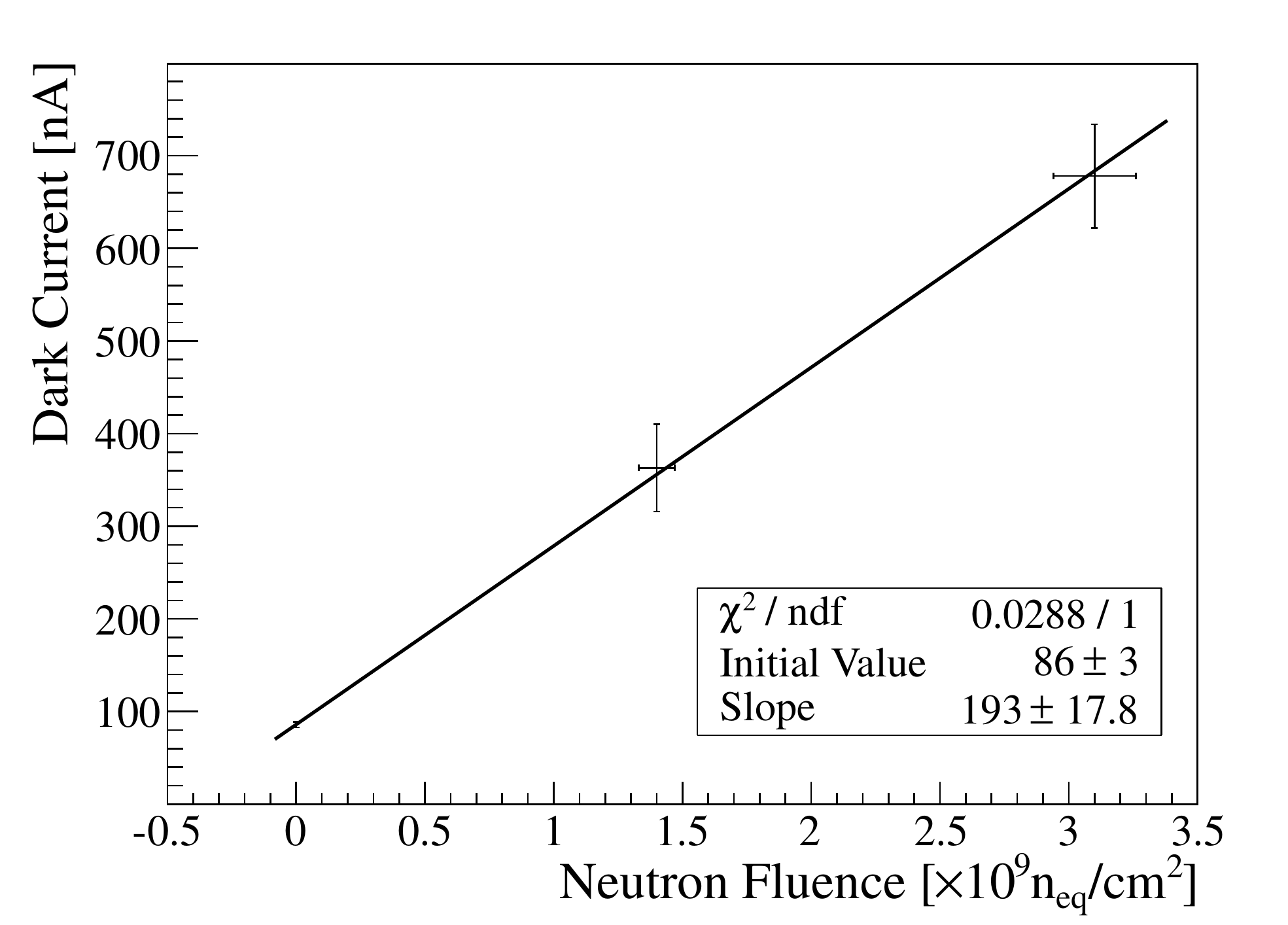}\end{center}
  \caption{Damage curve of 1~mm SiPM as a function of 1~MeV neutron fluence assuming annealing at 40-60$^{\circ}$C.
    The current was measured at 25$^{\circ}$C with a gain of 0.84$\times10^{6}$.}\label{fig:damage_curve}
\end{figure}
Finally, we plot the average dark current with annealing at temperature above 40$^{\circ}$C as a function of neutron fluence in Fig.~\ref{fig:damage_curve}.
The error bars represent the variation of dark current among units.
The slopes of the damage during the two irradiations are consistent and it is clear that the previous irradiation will not effect later ones.

\subsubsection{Relation between Dark Rate and Dark Current and Their Temperature Dependence}\label{sec:rate}

What was being measured all the time is the dark current, but it is the dark rate which actually affects the performance of the Barrel Calorimeter in Hall~D.
In order to measure the dark rate, a DAQ system using a gated ADC was set up to record the random dark pulses generated by SiPMs.
The length of the gate is chosen to be 200~ns which is longer than the pulse width of SiPMs, $\sim$80~ns from 10\% to 10\%.
Three $1\times1$~mm$^2$ SiPMs~\footnote{Their serial numbers are 1853, 1854 and 1855.}, were measured at various temperatures between $-5^\circ$C and 25$^\circ$C while the gain was kept constant by adjusting the bias voltage based on the characteristic curve provided by Hamamatsu`\cite{hamamatsu_sipm}.
In other words, the voltage setting above breakdown or over-bias was kept constant.
All three units were also part of the irradiation test therefore their ADC spectra after the irradiation were also taken for comparison.

\begin{figure}[h]
  \begin{center}\includegraphics[width=1.0\linewidth]{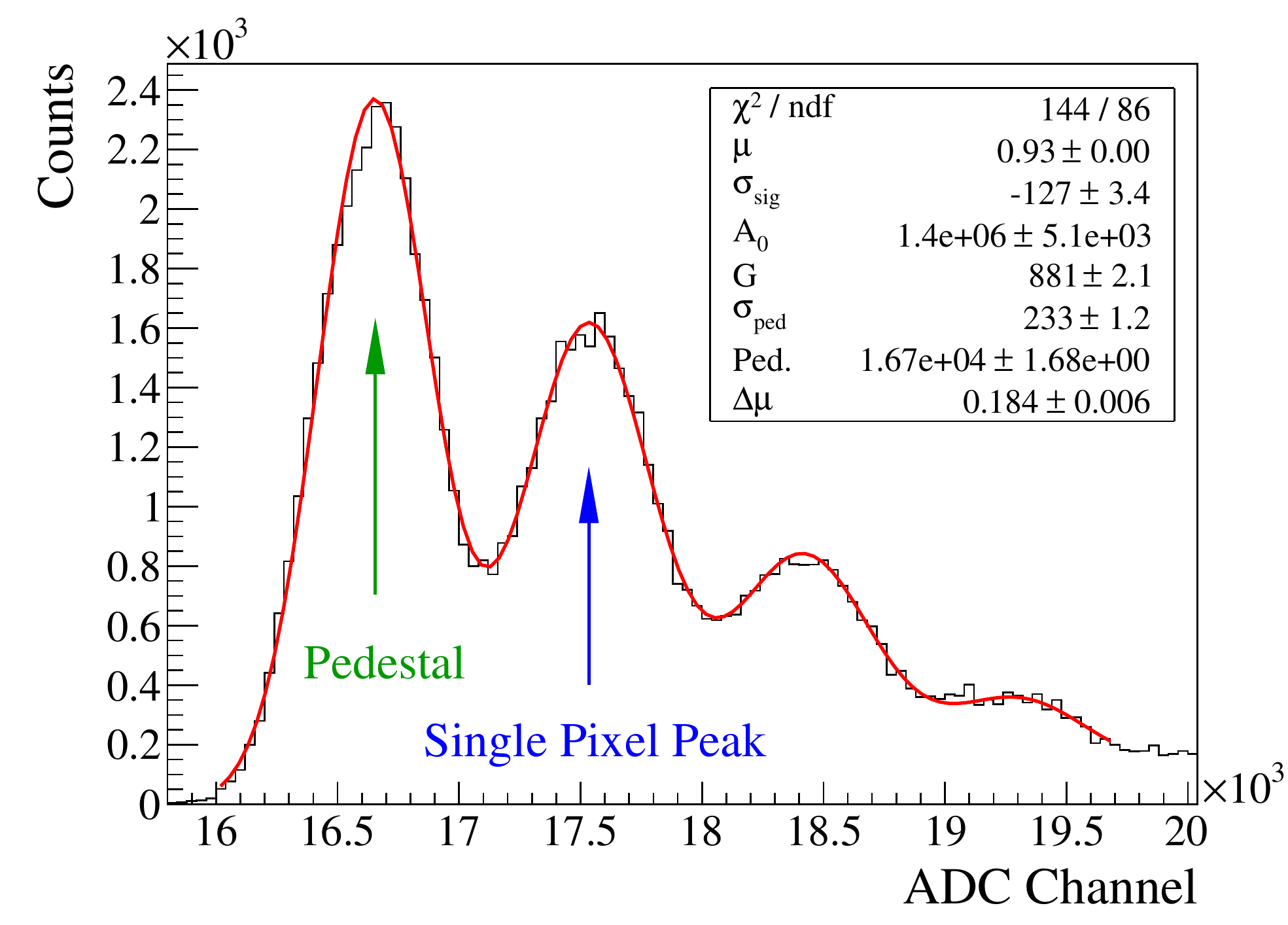}\end{center}
  \caption{A typical ADC spectrum for the dark rate measurement.
  The data were taken from SiPM \#1854 at 25$^\circ$C after the first irradiation.
  The description of the fitting function can be found in the text.}\label{fig:pde_fit}
\end{figure}
Fig.~\ref{fig:pde_fit} shows a typical ADC spectrum from a $1\times1$~mm$^2$ SiPM.
The histogram was fitted by a convolution of a discrete distribution function and a gaussian function.
\begin{equation}
    A(x) = A_0\cdot P(n|\mu,\Delta\mu)\otimes Gaus(x|n\cdot G + ped,\sigma_{n})\label{eqn:pde}
\end{equation}
where $A_0$ is the normalization factor.
The discrete distribution, $P(n|\mu,\Delta\mu)$, represents the probability that the number of pixels fired is equal to $n$, and it contains two Poisson distributions with one for the primary pixels fired and the other for the total of cross talk or after pulses~\cite{Renker2009} caused by the primary pixel:
\begin{eqnarray}
    P(n|\mu,\Delta\mu)&=&\sum_{n=i+j}{Pois(i|\mu)\cdot Pois(j|i\cdot\Delta\mu)}\nonumber\\
    &=&\sum_{n=i+j}{\frac{e^{-(\mu+i\Delta\mu)}\mu^{i}(i\Delta\mu)^j}{i!j!}}
\end{eqnarray}
where $\mu$ is the average number of primary pixels fired and $\Delta\mu$ is the average number of pixels fired around the primary pixel due to cross talk and after pulses.
The Gaussian function, $Gaus(x|n\cdot G + ped,\sigma_{n})$,represents the distribution of the charge with the number of pixels fired equal to $n$.
$G$ is the total gain~\footnote{It includes the intrinsic gain of an APD pixel, the gain of the pre-amplifiers and the ADC's analog-to-digital conversion factor.} in ADC channels and $ped$ is the ADC pedestal value.
The width, $\sigma_{n}$, is equal to
\begin{equation}
    \sigma_n = \sqrt{\sigma_\mathrm{ped}^2+n\cdot\sigma_\mathrm{sig}^2}
\end{equation}
where $\sigma_{ped}$ is the width of the pedestal and $\sigma_{sig}$ is the intrinsic width of a single pixel signal.
Such a function is valid when both the signal occupancy and the $\Delta\mu$ are small, which is true for our test condition.
Otherwise, the function needs to be modified by replacing the Poisson distributions with binomial distributions.
%Given our test condition, we were able to reliably extract the number of pixels fired in the 200~ns gate, the gain and the total probability of the cross talk and after pulses within the gate.

\begin{figure}[h]
  \begin{center}\includegraphics[width=1.0\linewidth]{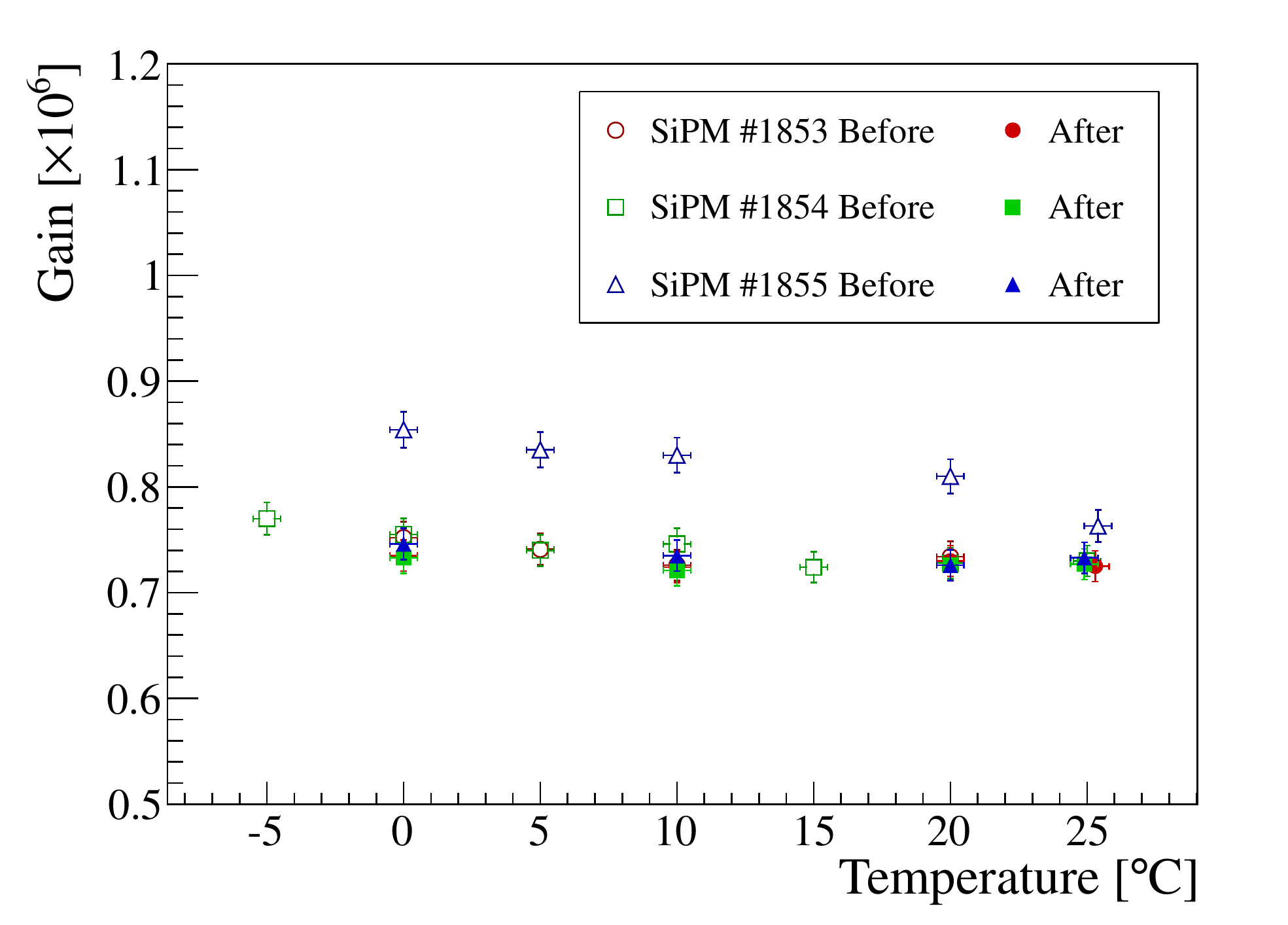}\end{center}
  \caption{The absolute gain of SiPMs at different temperatures at constant over-bias.
  Different legends represent the data from different units before and after the irradiation.}\label{fig:gain_temp}
\end{figure}
The absolute gain of SiPMs was calculated by dividing the total gain $G$ by the known ADC conversion factor and the gain of the preamplifiers.
As shown in Fig.~\ref{fig:gain_temp}, the gain is relatively stable at different temperatures as long as the bias voltage setting was adjusted to compensate for the change in the break down voltage with temperature to have a fixed over-bias.
The uncertainties shown in the plot only include the statistical uncertainty from the fit and the uncertainty of the temperature reading.
However, due to the temperature gradient in the cooling device, the gain fluctuates slightly around the average value, $0.84\times10^6$.
%Instead of fine tuning the bias voltage to get the exact gain expected, $7.5\times10^{5}$, the dark current at various temperatures was corrected afterwards by the ratio of the measured gain to the expected gain for comparison.

\begin{figure}[h]
  \begin{center}\includegraphics[width=1.0\linewidth]{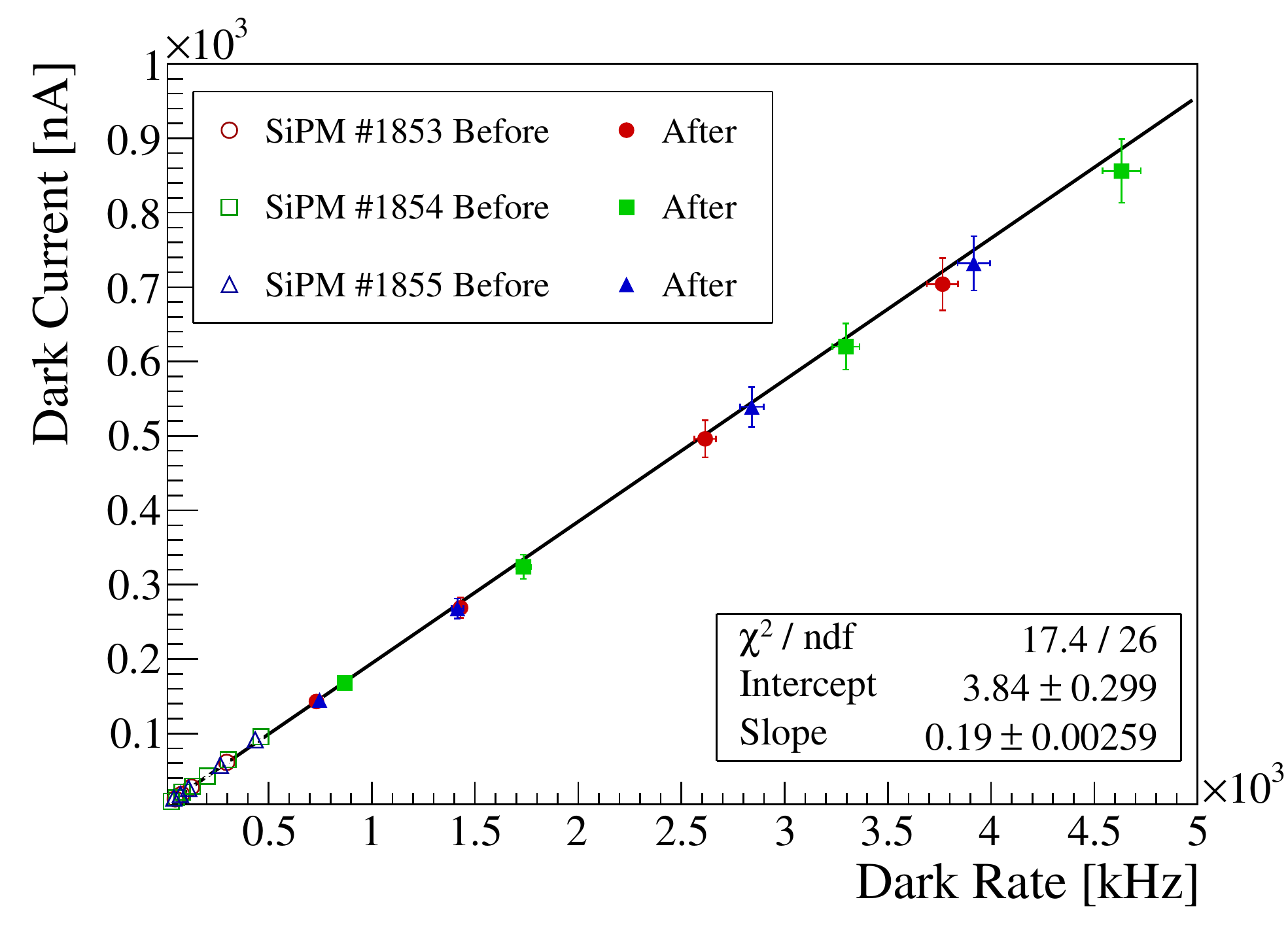}\end{center}
  \caption{The correlation between the dark current and dark rate with a gain of $0.84\times10^6$.
  }\label{fig:current_rate}
\end{figure}
Fig.~\ref{fig:current_rate} shows the correlation between the dark current and the dark rate, and the dark current has been corrected for the deviation of the gain to the average value.
The correlation is well described by a linear function:
\begin{equation}
    I = 0.190~\mathrm{nA/kHz}\cdot f + 3.84~\mathrm{nA}
\end{equation}
The slope corresponds to an average gain of $1.19\times10^{6}$ which is about 42\% higher than the actual gain.
Part of the mis-match comes from the cross talk and after pulses which are not counted in the extraction of the dark rate.
The rest may be attributed to the fact that Eqn.~(\ref{eqn:pde}) does not fully account for the pulses partially integrated in the ADC gate~\footnote{In retrospect, we realize that a longer gate width may have allowed a more accurate determination of the extracted dark rate from the fit.}.
Clearly, the radiation damage does not have any impact on this relation.

\begin{figure}[h]
  \begin{center}\includegraphics[width=1.0\linewidth]{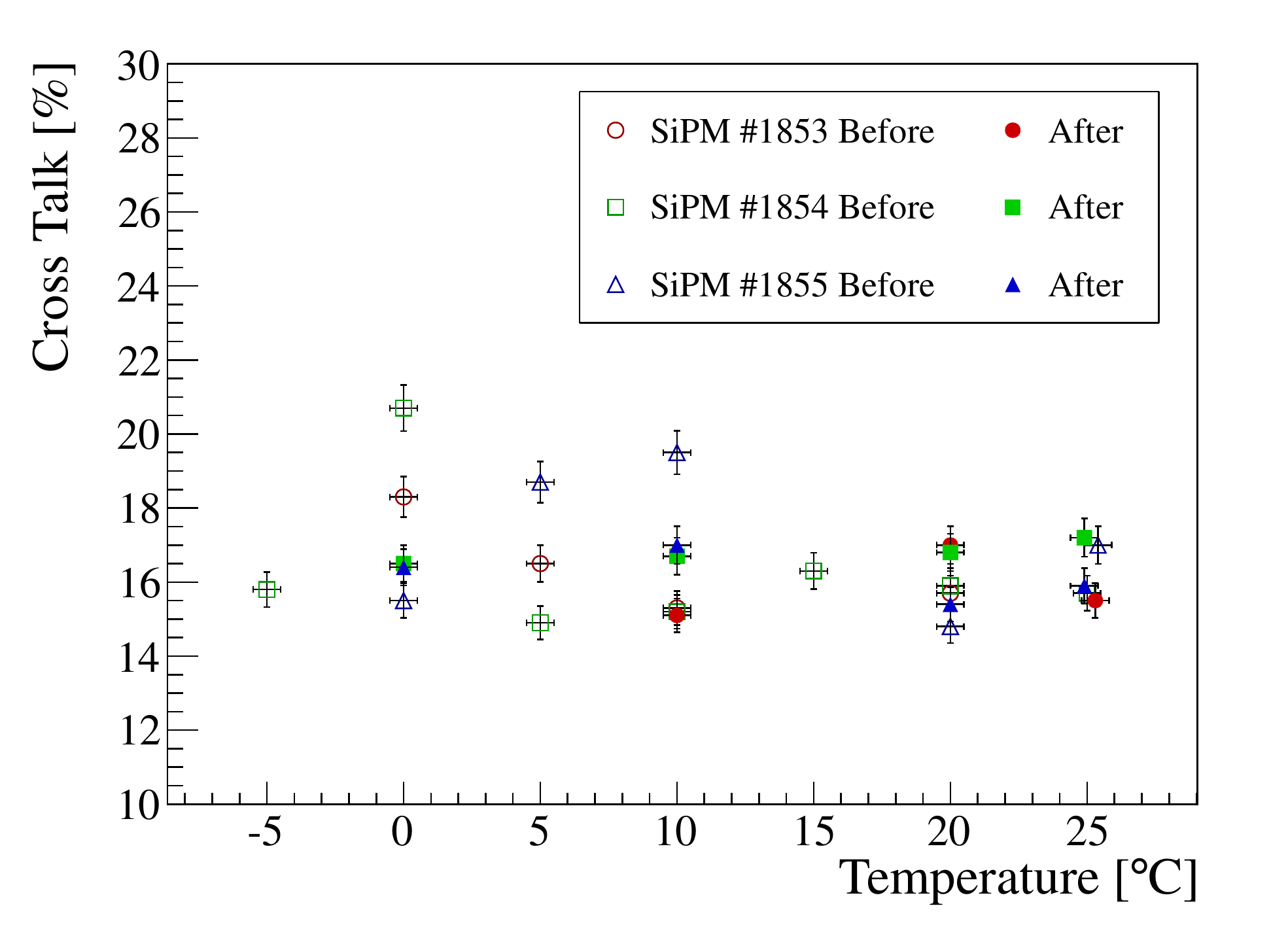}\end{center}
  \caption{The dependence of cross talk on temperature.}\label{fig:xtalk_temp}
\end{figure}
On the other hand, there is no significant temperature dependence of the cross talk and after pulses observed as shown in Fig.~\ref{fig:xtalk_temp}.
And the radiation damage doesn't change them as well.

\begin{figure}[h]
  \begin{center}\includegraphics[width=1.0\linewidth]{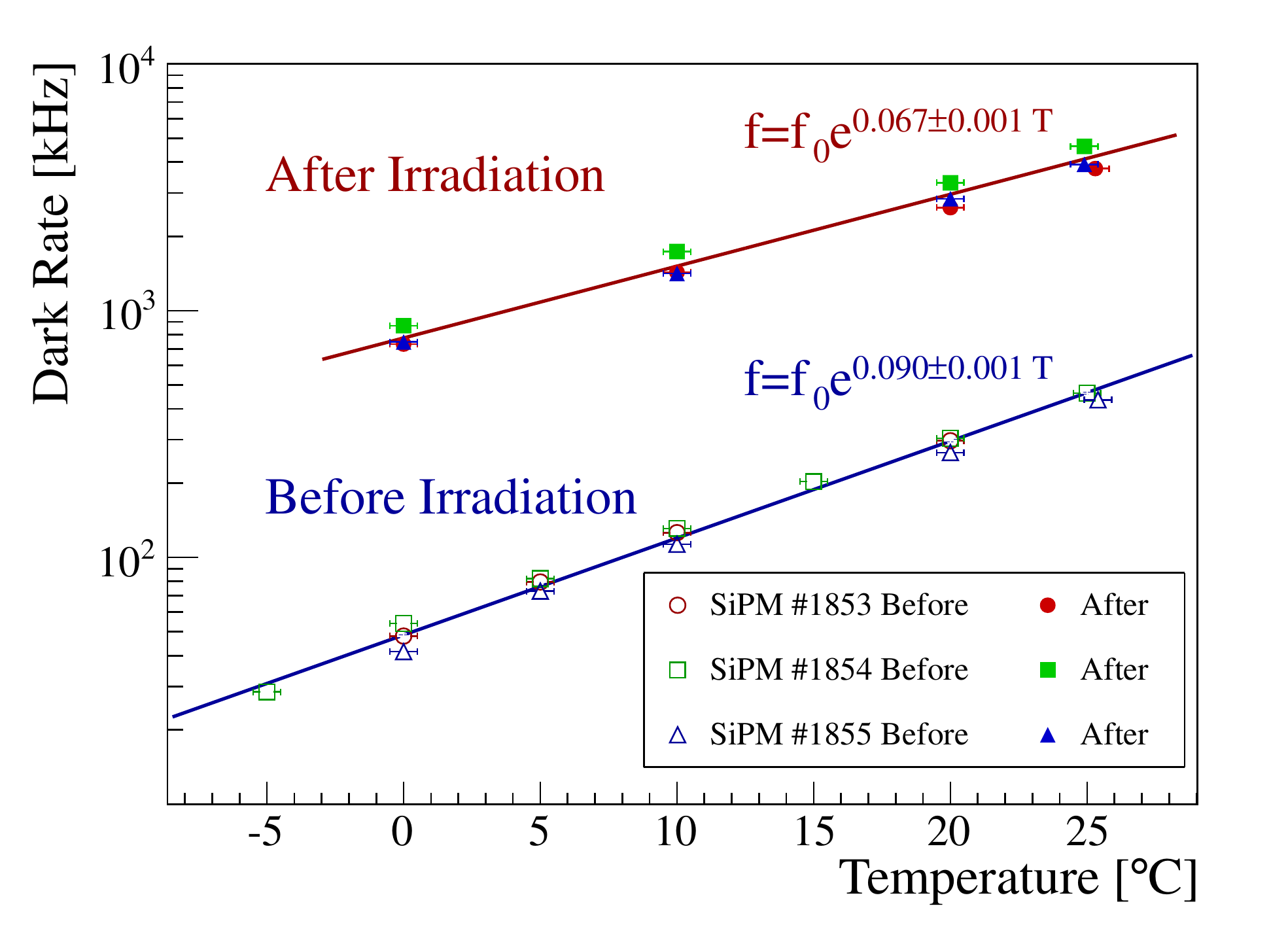}\end{center}
  \caption{The dependence of the dark rate before and after irradiation on temperature.}\label{fig:rate_temp}
\end{figure}
Fig.~\ref{fig:rate_temp} shows the temperature dependence of the dark rate before and after the neutron irradiation.
The dependence is exponential in the measured temperature range and the change of the temperature coefficient caused by the radiation damage is relatively small.
As a result, the average dependence is
\begin{equation}
    f = f_{0}\cdot e^{0.075\cdot(T-T_0)}
\end{equation}
Such a behavior provides the motivation to cool SiPMs during beam time in Hall~D to reduce the dark rate.

\section{Summary}

We measured the neutron radiation damage to SiPMs using neutrons generated by an electron beam at Jefferson Lab and a AmBe neutron source.
We further studied the temperature dependence of the radiation damage and other properties including dark rate, dark current and damage recovery.
We found that both dark rate and dark current increase linearly as a function of the total neutron fluence and the damage does not depend on the temperature or operating voltage.
Part of the acute damage will recover.
The speed and the extent of this annealing process strongly depends on the temperature and is faster and stronger at higher temperature.
Increasing the temperature of a damaged unit previously annealed at a certain temperature brings further recovery, but lowering the temperature will not reverse the recovery achieved.
We also measured the temperature dependence of the dark current and dark rate of SiPMs at a fixed gain.
Such a dependence is not strongly affected by the neutron radiation damage.

The results obtained by this study provided important information for implementing SiPMs as the readout of the Barrel Calorimeter in JLab Hall~D.



%% References

\section{Acknowledgement}

We acknowledge the JLab Hall~A staff for their support during the electron beam irradiation test.
We acknowledge M. Washington, J. Jefferson and D. Hamlette from the JLab RadCon group for providing AmBe source and calibrated neutron probes.
We also acknowledge Dr. P. Degtyarenko from the JLab RadCon group for calculating the neutron dose in Hall~A.
This work was supported by the U.S. Department of Energy. Jefferson Science Associates, LLC, operates Jefferson Lab for the U.S. DOE under U.S. DOE contract DE-AC05-060R23177.


\bibliographystyle{elsarticle-num}
\bibliography{SiPM_HallD}






% \end{thebibliography}

\end{document}